\documentclass{article}

\usepackage{arxiv}

\usepackage[utf8]{inputenc} 
\usepackage[T1]{fontenc}    
\usepackage{url}            
\usepackage{booktabs}       
\usepackage{amsfonts}       
\usepackage{nicefrac}       
\usepackage{microtype}      
\usepackage{graphicx}

\title{A loop enhancement strategy for network robustness}

\author{
  Masaki Chujyo\thanks{Corresponding author} \\
  Japan Advanced Institute of Science and Technology\\
  Ishikawa, 923-1292 Japan\\
  \texttt{mchujyo@jaist.ac.jp} \\
   \And
 Yukio Hayashi \\
  Japan Advanced Institute of Science and Technology\\
  Ishikawa, 923-1292 Japan\\
  \texttt{yhayashi@jaist.ac.jp} \\
}

\begin{document}
\maketitle

\begin{abstract}
Many real systems are extremely vulnerable against attacks, since they are scale-free networks as commonly existing topological structure in them.
Thus, in order to improve the robustness of connectivity, several edge rewiring methods have been so far proposed by enhancing degree-degree correlations. 
In fact, onion-like structures with positive degree-degree correlations are optimally robust against attacks. 
On the other hand, recent studies suggest that the robustness and loops are strongly related to each other. 
Therefore, we focus on enhancing loops as a new approach for improving the robustness. 
In this work, we propose edge rewiring methods and evaluate the effect on the robustness by applying to real networks. 
Our proposed methods are two types of rewirings in preserving degrees or not for investigating the effect of the degree modification on the robustness. 
Numerical results show that our proposed methods improve the robustness to the level as same or more than the state-of-the-art methods. 
Furthermore, our work shows that the following two points are more important for further improving the robustness. 
First, the robustness is strongly related to loops more than degree-degree correlations. 
Second, it significantly improves the robustness by reducing the gap between the maximum and minimum degrees.
\end{abstract}


\section{Introduction}
Improving robustness against malicious attacks has been one of the important issues in network science. 
Because many real networks are scale-free whose degree distributions follow power-law, and their connectivity is extremely vulnerable against removal of targeted nodes \cite{barabasi1999emergence,albert2000error,cohen2001breakdown}. 
This vulnerability causes the loss of essential functions as networks in many real systems, which operate on the assumption that all nodes are connected.

For example, the following infrastructures were damaged by disasters, and caused significant impacts on our society. 
A massive ice storm in Eastern Canada caused a long-term blackout in 1998 \cite{chang2007infrastructure}. 
It reveals the strong dependence on electric power. 
Also, a power outage in the USA and Canada continued for two days in 2003, and caused transportation and economic disruptions \cite{minkel20082003}. 
In 2008, a power grid network in China broken down due to heavy snowfall \cite{zhou2016two}. 
On the other type of disasters, the eruption of Icelandic volcano Eyjafjallaj\"{o}kull in 2010 affected European air traffic and stranded thousands of passengers \cite{brooker2010fear}.
The earthquakes and tsunamis that struck Japan in 2011 caused a terrible loss of life and property further disrupting the global supply chain network \cite{mackenzie2012measuring}.
In 2012, Hurricane Sandy destroyed the large areas in New York and New Jersey \cite{manuel2013long}. 
After that, the blackout in several months affected the transportation network, and caused multiple damages \cite{lipton2013cost}. 
Thus, these infrastructures directly connected to our life have potential risks, and it is necessary to design a new structure to mitigate the outage.

They are also seriously damaged by attacks with targeting a specific part.
For example, the North American power network is robust against failures, but its function is greatly reduced for targeted attacks \cite{albert2004structural}.
Moreover, an assessment of the urban rail transport network indicates that the Shanghai Metro is vulnerable to degree-based attacks \cite{sun2015vulnerability}.
In investigating the robustness of the global air transport network against intentional attacks, the weak points are discussed in a viewpoint from each airport's centrality \cite{lordan2014robustness}.
Furthermore, the robustness is analyzed against both failures and attacks of airline network routes in combining Low Cost Carriers (LCCs) and Full Service Carriers (FSCs) \cite{lordan2016robustness}.
It is concluded that route networks of LCC are more robust than ones of FSC.

Since many of the above infrastructures in daily life have vulnerability against attacks, several methods should be developed for improving the robustness of connectivity in network systems.
We remark that such systems can be more robust by partial edge rewiring without adding new resources of edges \cite{schneider2011mitigation}.
Thus, we aim to improve the robustness without adding any resources in assuming that the number of both nodes and edges is constant.
Particularly, the edges in the airline network or the wireless communication network can be easily changed as rewiring. 
The rewiring with preserving degrees in the airport network or wireless communication is possible by changing the destination of the airport or the direction of the wireless beam.
In contrast, it may be difficult to change the network structure when edges are spatially embedded, such as road networks, water supply networks, and power grid networks.
However, even on such systems, it will be useful for maintaining network functions to make the robust structure by adding new resources or renovating and rebuilding.

Although we discuss the robustness of connectivity by attacks, we may consider other attacks. 
Some edge rewiring algorithms have also been proposed as adversarial attacks against link prediction \cite{SYu2019Target} and community detection \cite{jchen2019GABase}.
While they aim to rewire the connections for privacy protections, we aim to enhance the tolerance of connectivity against node removals.

On the other hand, an onion-like structure with positive degree-degree correlations \cite{newman2002assortative} is optimally robust against targeted attacks under a given degree distribution \cite{schneider2011mitigation,tanizawa2012robustness}.
The degree-degree correlations $r$ is defined as the Pearson correlation coefficient for degrees at both ends of an edge \cite{newman2002assortative}.
The onion-like structure is visualized by arranging similar degree nodes on a concentric circle in decreasing order of degrees from the core to the peripheral.
Since similar degree nodes tend to be connected by the positive degree-degree correlations, they draw a circle.
It can be generated by greedy rewiring to maximize a robustness index $R_\mathrm{hub}$, which accumulates the size of the largest connected component after attacks \cite{schneider2011mitigation}. 
The robustness index $R_\mathrm{hub}$ is defined as $R_\mathrm{hub}=\frac{1}{N}\sum_{q={1/N}}^{1}S(q)$, 
where $S(q)$ denotes the number of nodes included in the largest connected component after removing $qN$ nodes, and $q$ is the fraction of removal nodes by high degree adaptive attacks.
However, there is no strict definition of an onion-like structure for the thresholds of $R_\mathrm{hub}$ or $r$, since too high degree-degree correlations rather decrease the robustness \cite{schneider2011mitigation,tanizawa2012robustness,murakami2017robustness}.
Thus, through numerical simulations, it is considered that the onion-like networks have $R_\mathrm{hub}>0.3$ and $r>0.2$ because not onion-like scale-free networks by Barab\'{a}si-Albert model \cite{barabasi1999emergence} have $R_\mathrm{hub}<0.23$ and $r\approx 0$ at the same size \cite{hayashi2018onion}.
The values of $R_\mathrm{hub}$ and $r$ for onion-like networks are also obtained by rewiring for enhancing degree-degree correlations \cite{wu2011onion}.
Note that the $R_\mathrm{hub}$ represents the area under the curve $S(q)/N$ versus $q=\frac{1}{N}, \frac{2}{N},...\frac{N-1}{N}, 1$, and can take different values for networks with the same critical point $q_c$ for whole fragmentation. 
In other words, the $R_\mathrm{hub}$ takes a large value when $S(q)$ decreases steeply at $q_c$, whereas it takes a small value when $S(q)$ decreases gently even at the same $q_c$.

Based on enhancing the degree-degree correlations, several rewiring methods have been proposed for improving the robustness \cite{xulvi2002evolving,wu2011onion}.
However, in recent years, an incrementally growing method is also proposed for constructing an onion-like network by enhancing loops (or cycles in graph theory) instead of the degree-degree correlations \cite{hayashi2018new,hayashi2018onion}. 
It has been suggested that there is a strong relation between robustness and loop structure.

In this work, we propose new rewiring methods to enhance loops, and discuss the topological structures in improving the robustness for real data of the infrastructure networks.
We emphasize the relation between robustness and loop rather than the conventional degree-degree correlations.

\section{Methods}
We explain our motivations for the rewiring strategy in enhancing loops. 
Several methods have been so far proposed for improving the robustness to be an onion-like structure by increasing the degree-degree correlations \cite{xulvi2002evolving,wu2011onion}.
However, a network with the extremely high degree-degree correlations is not the best \cite{schneider2011mitigation,tanizawa2012robustness,murakami2017robustness}. 
Therefore, for the improvement of robustness, there may exist other approaches instead of the degree-degree correlations.

\begin{figure}[ht!]
  \centering
  \includegraphics[width=.7\linewidth]{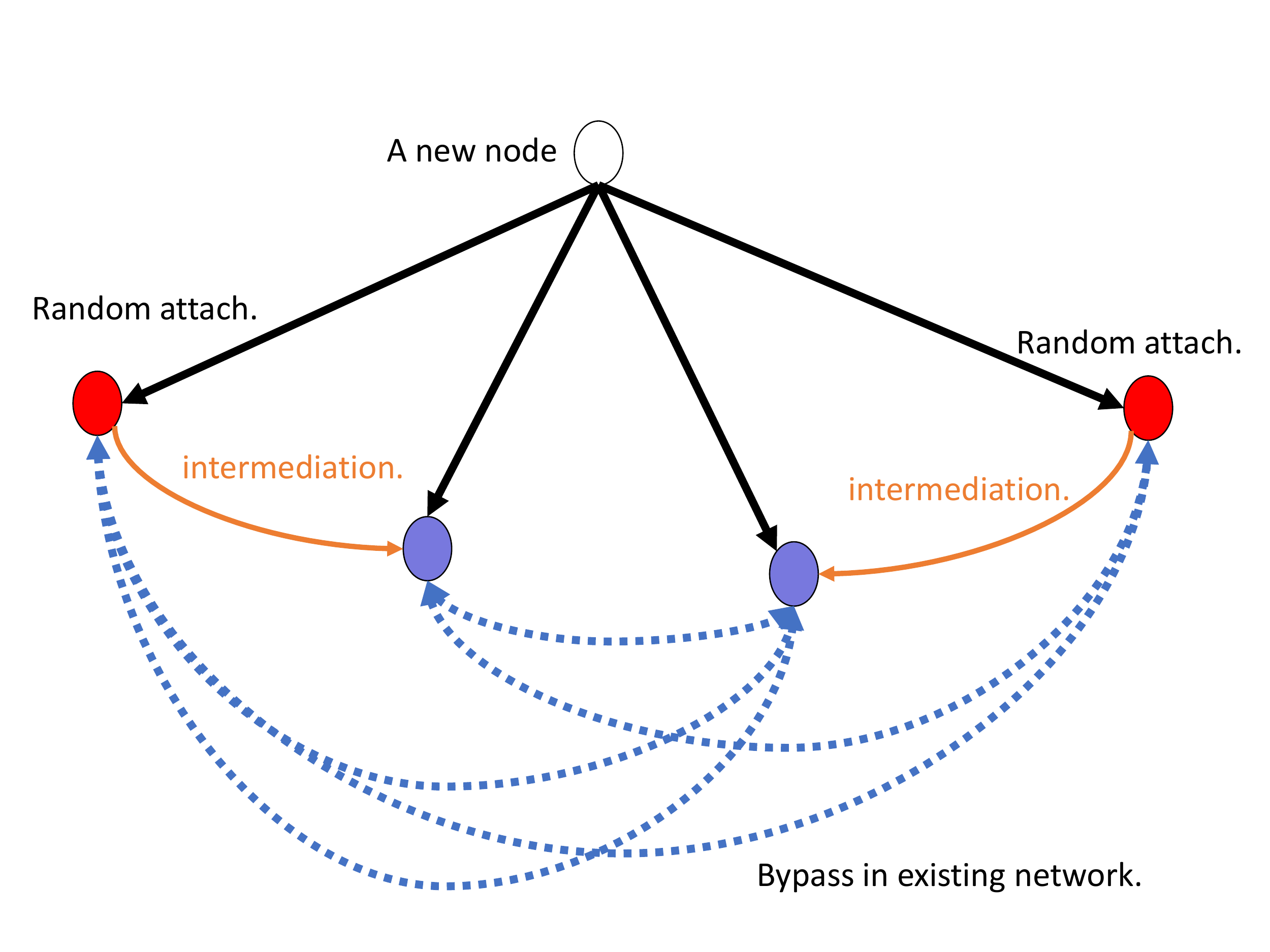}
  \caption{Illustration of pairs of intermediation attachments.
      Black bold and blue dashed lines denote added edges and existing paths in the network. Orange line denote paths of intermediations from randomly selected nodes. }
      \label{MED-attach}
\end{figure}

We remark a strong relation of robustness and loops from some suggesting works \cite{braunstein2016network,hayashi2018onion}.
One of them is the equivalence of network dismantling and decycling problems \cite{braunstein2016network}.
Here, network dismantling problem is finding a minimum set of nodes that removal makes the network broken into connected components at most a given size.
Network decycling problem is finding a minimum set of nodes that removal makes the network without loops.
The decycling set is named as Feedback Vertex Set (FVS) in computer science.
The equivalence means that networks become a tree structure at the critical point before the whole fragmentation.
Therefore, in order to avoid fragmentation, it is necessary not to be a tree as long as possible against node removals.
On the other hand, the relation of the robustness and loops is also discussed in generating the onion-like structure with the optimal tolerance of connectivity against attacks. 
In the generation based on a pair of random and intermediation attachments, a new node links to a randomly selected node and the minimum degree node in distant neighbors through a few hops of intermediation from the randomly selected pair node \cite{hayashi2018onion}.
We intuitively understand that many loops with bypasses are formed by pairs of attachments as shown in Fig. \ref{MED-attach}.
In fact, it is found that the robustness index $R_{\mathrm{hub}}$ and the size of FVS have strong correlations in the networks.

Focusing the above correlations, we propose a new edge rewiring strategy for enhancing loops in two types: Preserving and Non-Preserving with modification of the degree distribution.
As similar to the conventional methods, in Preserving, the rewiring does not change each node's degree under the original degree distribution.
However, in Non-Preserving, the rewiring changes degrees and the degree distribution in order to investigate the effect of changes in the degree distribution on robustness.

\subsection*{Spanning trees and the fundamental cycles}
The fundamental system of cycles (loops) in a spanning tree is known in graph theory. 
A spanning tree is a subgraph that all nodes of the network are connected without loops. 
The chords are edges not belonging in the spanning tree.
Each chord and a loop called a fundamental cycle are one-to-one correspondings \cite{bollobas2013modern}. 
In other words, a spanning tree has $M-N+1$ fundamental cycles as a linearly independent basis, where $M$ and $N$ denote the numbers of edges and nodes. 
Therefore, any loop is represented as a combination of the basis.
It is expected that there are many loops independently on networks with a large number of spanning trees.
Independently from the above our explanation, a rewiring method for increasing the number of spanning trees has been proposed by applying the perturbation theory of the Laplacian matrix \cite{chan2016optimizing}.
The authors consider a rewiring in Preserving by the addition and removal of edges based on the Kirchhoff’s matrix-tree theorem \cite{BUEKENHOUT199869}.
In contrast, we consider the edge rewiring strategy to enhance loops by increasing the size of Feedback Vertex Set instead of the number of spanning trees. 

\subsection*{Feedback Vertex Set}
Since the Feedback Vertex Set (FVS) is the minimum set of nodes whose removal makes the network acyclic, the remaining trees after removing the FVS are easily fragmented by further removal.
The attack method to a node estimated in the FVS by Belief Propagation is proposed \cite{mugisha2016identifying}.
Inversely, it is expected that increasing the size of FVS leads to improve the robustness of connectivity against node removal.

For increasing the size of FVS, we propose a new rewiring strategy by enhancing loops to improve the robustness. 
Since to find the FVS belongs to a class of NP-hard combinatorial optimization problems, the exact solution is intractable for a large network. 
Therefore, we apply an approximation algorithm of Belief Propagation (BP) in statistical physics \cite{zhou2013spin}.
The algorithm estimates the probability $q_i^0$ belonging to FVS for node $i$.
Here, $q_i^{A_i}$ denotes the marginal probability for node $i$'s root: $A_i=0$ (empty) or $A_i=i$ (the root is itself).
When the node is empty, it is unnecessary as a root so that it is estimated as belonging to the FVS.
Based on a cavity method \cite{zhou2013spin,mugisha2016identifying}, the explicit formulas are 
\begin{equation}
    q^0_i = \frac{1}{z_i},
\end{equation}
\begin{equation}
    z_i=1+e^x\left[ 1+\sum_{k\in \partial i}\frac{1-q_k^0}{q_{k\rightarrow i}^0+q_{k\rightarrow i}^k}\right]\prod_{j\in \partial i}[q_{j\rightarrow i}^0+q_{j\rightarrow i}^j],
\end{equation}
where $\partial i$ denotes node $i$'s set of neighbor nodes, $x>0$ is a parameter of inverse temperature, and $z_i$ is normalization constant.
The $q_{i\rightarrow j}^0$ and $q_{i\rightarrow j}^i$ are calculated from the following self-consistent BP equations,
\begin{equation}
    q_{i\rightarrow j}^0=\frac{1}{z_{i\rightarrow j}},
\end{equation}
\begin{equation}
    q_{i\rightarrow j}^i=\frac{e^x\prod_{k\in \partial i\backslash j}[q_{k\rightarrow i}^0+q_{k\rightarrow i}^k]}{z_{i\rightarrow j}},
\end{equation}
\begin{equation}
    z_{i\rightarrow j}=1+e^x\prod_{k\in \partial i\backslash j}[q_{k\rightarrow i}^0+q_{k\rightarrow i}^k]
    \times \left[1+\sum_{l\in \partial i\backslash j}\frac{1-q_{l\rightarrow i}^0}{q_{l\rightarrow i}^0+q_{l\rightarrow i}^l}\right],
\end{equation}
where $\partial i\backslash j$ denotes node $i$'s set of neighbor nodes except node $j$, and $z_{i\rightarrow j}$ is normalization constant. 
Equations (1)-(5) are iterated from an initial set of random values in $(0,1)$ until given rounds in practically.
In each round, a set $\{q_i^0 | i=1,...,N\}$ are updated in order of random permutation of the all $N$ nodes.
To obtain the FVS, we remove a node $i$ with a higher $q_i^0$ and recalculate a set $\{q_i^0\}$ for all existing nodes until given rounds.
The removed nodes are estimated as the FVS.
We repeat them until the network without loops.
A node $i$ with a smaller $q^0_i$ is less related to loops.
In other words, a node $i$ with a smaller $q^0_i$ tends not to belong to FVS.
Using this $\{q^0_i\}$, we consider the edge addition and deletion for increasing the size of FVS as follows.

\begin{figure}[h!]
  \centering
  \includegraphics[width=.7\linewidth]{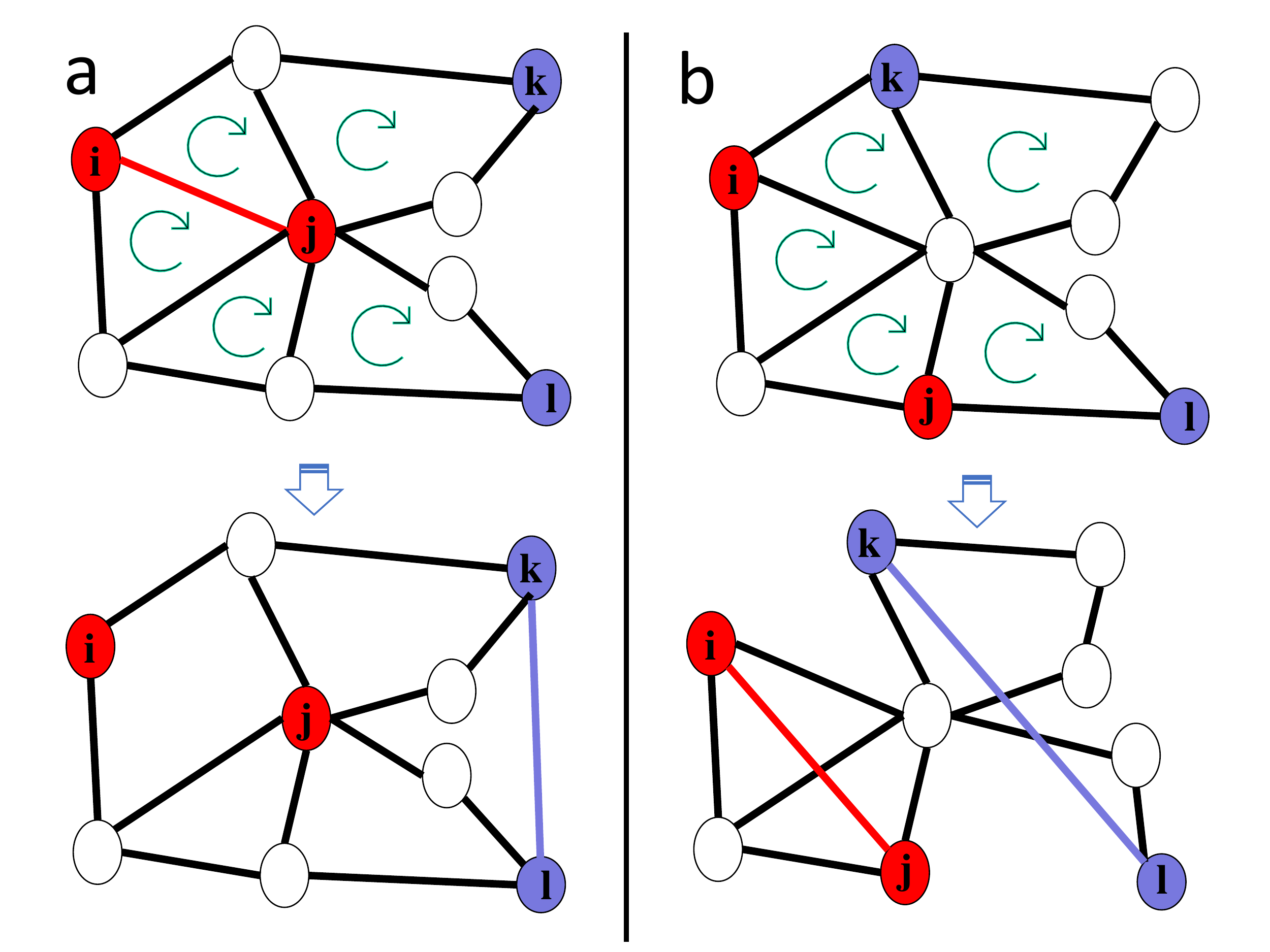}
  \caption{Illustration of our proposed methods.
      (a) BP Non-Preserving, (b) BP Preserving. The nodes with larger $q^0_i$ and $q^0_j$ and the edges between them are filled with red, while the nodes with smaller $q^0_k$ and $q^0_l$ and the edges between them are filled with blue.}
      \label{illust}
\end{figure}

\subsection*{BP Non-Preserving}
The following rewiring without preserving degrees is called BP Non-Preserving. 
Here this rewiring modifies degrees under a constant number of nodes and edges.
In BP Non-Preserving, to increase the size of FVS, we add an edge $(k,l)$ between nodes with smaller $q^0_k$ and $q^0_l$ in the all unconnected nodes pairs. 
It is expected to increase the size of FVS by adding the edge, since these nodes tend not to belong to any loops, and their connection makes a new loop.
To keep the number of edges, we remove an edge $(i,j)$ between nodes with larger $q^0_i$ and $q^0_j$ in the all connected nodes pairs.
Removing the edge $(i,j)$ has little impact on the size of FVS, since the nodes $i$ and $j$ are on many loops because of large $q_i^0$ and $q_j^0$ as candidates of FVS.
As shown in Fig. \ref{illust}a, the following steps are repeated in BP Non-Preserving.
Note that we may exchange steps 1 and 2 because they are independent processes.
\begin{description}
    \item[Step 1.] Add a non-existing edge $(k,l)$ with the minimum $q^0_k$ and $q^0_l$.
    \item[Step 2.] Remove an edge $(i,j)$ with the maximum $q^0_i$ and $q^0_j$.
    \item[Step 3.] Recalculate $\{q^0_i | i=1,...,N\}$.
\end{description}

\subsection*{BP Preserving}
The following rewiring with preserving degrees is called BP Preserving.
As similar to \cite{chan2016optimizing}, we apply the three steps:({\sl i}) add a non-existing edge $(i,j)$, ({\sl ii}) remove edges $(i,k)$ and $(j,l)$, and ({\sl iii}) add a non-existing edge $(k,l)$. 
We illustrate the steps before and after rewiring at the top and bottom in Fig. \ref{illust}b, respectively.
It consists of the additions of two edges $(i,j)$ and $(k,l)$, and the removal of two edges $(i,k)$ and $(j,l)$ in order to preserve degrees.

For increasing the size of FVS, it is effective to add two edges between nodes with smaller $q^0_i$.
However, it can not apply to BP Preserving, since the removal for preserving degrees in the above step ({\sl ii}) makes fragmentation.
Nodes with smaller $q_i^0$ tend to have smaller degrees and belong to a tree.
In the worst-case, when we select nodes $i$ and $j$ with degree one or on a dangling tree, they are isolated by the removal.
To avoid it, we select two unconnected nodes $i$ and $j$ with larger $q_i^0$ and $q_j^0$ in all nodes at first, since they tend to have large degrees and removing edges emanated from them is not likely to decrease the connectivity.
Then, we select unconnected nodes $k$ and $l$ with smaller $q_k^0$ and $q_l^0$ in the neighbors of nodes $i$ and $j$, respectively, in order to enhance loops by connecting them in the above step ({\sl iii}).
Since the nodes $k$ and $l$ are selected in the neighbors of nodes $i$ or $j$, not in all nodes, they may be a little contained in loops, which is not the worst-case and prevents isolation.
In this way, we first add an edge $(i,j)$ between nodes $i$ and $j$ with larger $q_i^0$ and $q_j^0$ in avoiding fragmentation by rewiring as much as possible.
These selections possibly increase the size of FVS, since the nodes $k$ and $l$ with smaller $q_k^0$ and $q_l^0$ are expected to be included in new loops.
For the removal, we select edges $(i,k)$ between nodes with larger $q^0_i$ and smaller $q^0_k$, and $(j,l)$ between nodes with larger $q^0_j$ and smaller $q^0_l$.
The removals have little impact on the size of FVS, since the edges linked to nodes with smaller $q^0_k$ or $q^0_l$ tend to belong to fewer loops.
As shown in Fig. \ref{illust}b, the following steps are repeated in BP Preserving.
\begin{description}
    \item[Step 1.] Let $(i, j)$ be a non-existing edge with the maximum $q^0_i$ and $q^0_j$ in a network.
    \item[Step 2.] Let $k$ be a node with the minimum $q^0_k$ in the neighbor of either node $i$. Let $l$ be a node with the minimum $q^0_l$ in the neighbor of node $j$ but not the neighbor of node $k$.
    \item[Step 3.] Add non-existing edges ($i$,$j$) and ($k$,$l$) and remove edges ($i$,$k$) and ($j$,$l$).
    \item[Step 4.] Recalculate $\{q^0_i | i=1,...,N\}$.
\end{description}

\begin{figure}[h!]
  \centering
  \includegraphics[width=.9\linewidth]{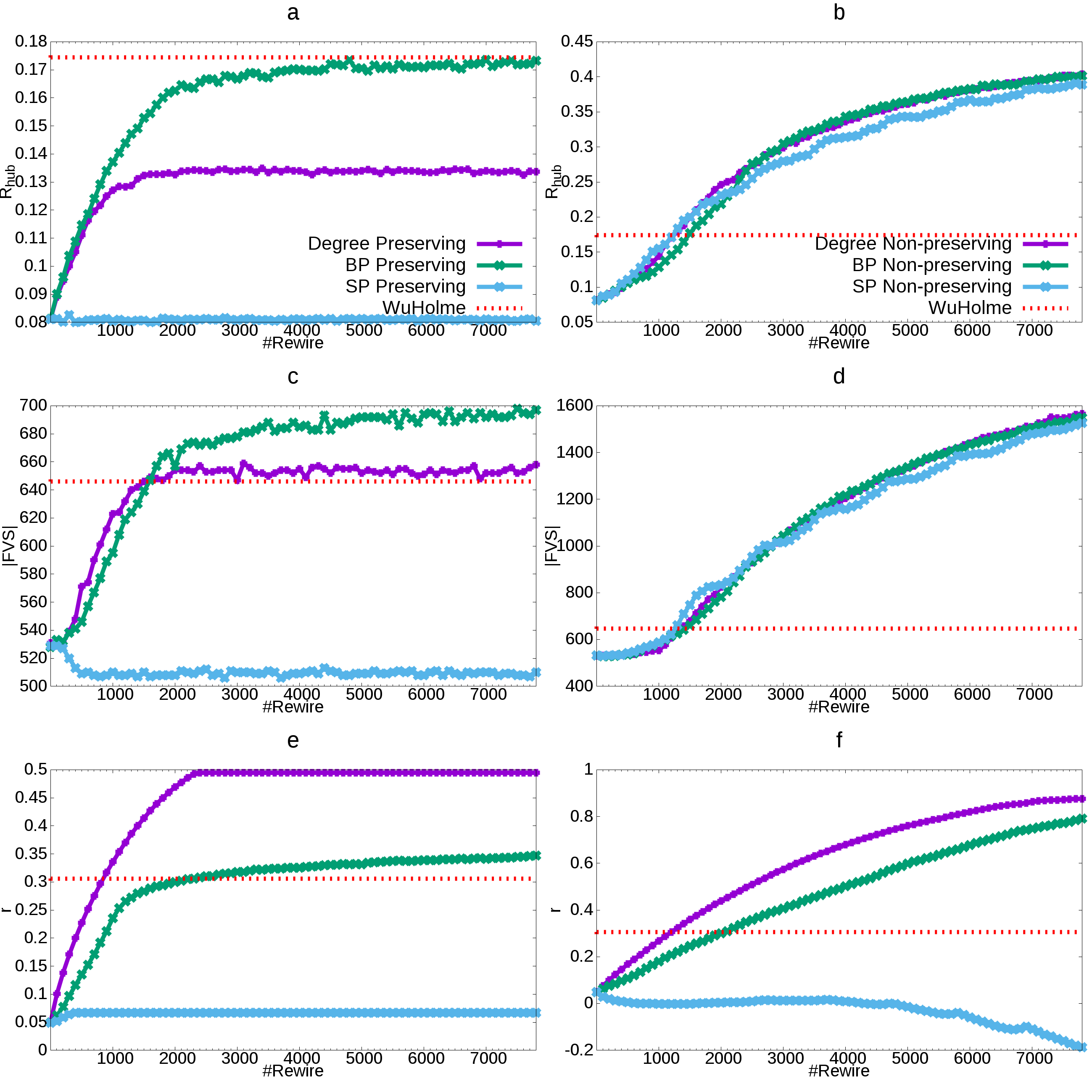}
  \caption{The robustness index, the approximate size of FVS, and the degree-degree correlations vs. the number of rewiring.
    (Left: a, c, e) Rewirings in Preserving. Violet, green, and light blue solid lines denote the result by Degree, BP, and SP Preserving, respectively. The red dot line indicates a baseline of the conventional best. (Right: b, d, f) Rewirings in Non-Preserving. Violet, green, and light blue solid lines denote the result by Degree, BP, and SP Non-Preserving, respectively.}
    \label{flog2}
\end{figure}

\section{Results}
In this section, we evaluate the effects of enhancing loops on the improvement of robustness in our proposed methods.
Also, we discuss a relation between the robustness and the size of FVS.
For comparison, Degree, WuHolme \cite{wu2011onion}, and SP \cite{chan2016optimizing} are investigated. 
Degree is a modification of our rewiring strategy between nodes with smaller degrees instead of smaller $q^0_i$ of BP.
It has two types: Degree Preserving and Degree Non-Preserving, corresponding to BP Preserving and BP Non-Preserving. 
We consider WuHolme as a baseline in Preserving, because it is the best conventional method for improving the robustness by increasing the degree-degree correlations. 
SP is the previously mentioned method for increasing the number of spanning trees, and it also has two types: SP Preserving and SP Non-Preserving.
We compare it as a different approach in order to enhance loops.

We apply our proposed and conventional methods for several real networks including social, biological, and technological networks, for which similar results are obtained (See Additional File).
As a typical result, Figure \ref{flog2} shows the robustness index $R_\mathrm{hub}$, the approximate size of FVS, and the degree-degree correlations $r$ versus the number of rewirings for a real network: an airline network named OpenFlights with $N = 2905$ nodes and $M = 15,645$ edges \cite{kunegis2013konect}.
The robustness index $R_\mathrm{hub}$ is the sum of the fraction of nodes in the largest connected component against high degree adaptive attacks with recalculation of degrees \cite{schneider2011mitigation}.
The degree-degree correlations $r$ is the Pearson correlation coefficient for degrees \cite{newman2002assortative}.
In the following, $|$FVS$|$ denotes the size of FVS by Belief Propagation \cite{zhou2013spin}, and $\mathrm{\#Rewire}$ is the number of rewirings.
Note that almost all edges are rewired at $\mathrm{\#Rewire}=7800$, which is nearly half of $M$.

First, we show the results for our proposed BP Preserving.
At the left in Fig. \ref{flog2}c, BP Preserving (denoted by the green line) increases up to $|\mathrm{FVS}|=698$ from the original $|\mathrm{FVS}|=528$ before rewiring at $\mathrm{\#Rewire}=0$. 
The rate of FVS rises from 18\% to 24\%. 
Since the baseline (denoted by the red dot line) is $|\mathrm{FVS}| = 646$, BP Preserving increases $|\mathrm{FVS}|$ over the baseline.
Furthermore, at the left in Fig. \ref{flog2}a, BP Preserving (denoted by the green line) improves $R_\mathrm{hub}$ to almost the same as the baseline (denoted by the red dot line).
The maximum $R_\mathrm{hub}$ is 0.175 in BP Preserving (denoted by the green line) and 0.174 in the baseline (denoted by the red dot line) at the left in Fig. \ref{flog2}a.
The maximum $R_\mathrm{hub}$ in Degree Preserving (denoted by the violet line) is 0.136, which is smaller than that in BP Preserving (denoted by the green line).
Therefore, BP Preserving increases $|$FVS$|$ more than other methods of Degree and SP Preserving and improves $R_\mathrm{hub}$ to almost the same level as the conventional best in Preserving. 
Enhancing the loop effectively improves the robustness.

Next, we show the results in BP Non-Preserving. 
At the right in Fig. \ref{flog2}d, BP Non-Preserving (denoted by the green line) increases to the maximum $|\mathrm{FVS}| = 1549$, which is 53\% of nodes are included in FVS.
It is about twice larger than $|\mathrm{FVS}|=698$ for the baseline (denoted by the red dot line) at the right in Fig. \ref{flog2}d.
In addition, other rewirings in Non-Preserving also increase $|$FVS$|$ at the same level as BP Non-Preserving.
The maximum $|$FVS$|$ is 1566 in Degree Non-Preserving (denoted by the violet line) and 1527 in SP Non-Preserving (denoted by the light blue line) at the right in Fig. \ref{flog2}d. 
At the right in Fig. \ref{flog2}b, BP Non-Preserving (denoted by the green line) increases to the maximum $R_\mathrm{hub} = 0.404$.
It is also about twice larger than $R_{\mathrm{hub}}=0.174$ for the baseline (denoted by the red dot line).
As similar to the result for $|$FVS$|$, other methods also increase $R_\mathrm{hub}$ to almost the same level as BP Non-Preserving.
The maximum $R_\mathrm{hub}$ is 0.405 in Degree Non-Preserving (denoted by the violet line) and 0.392 in SP Non-Preserving (denoted by the light blue line) at the right in Fig. \ref{flog2}b.
Therefore, in comparison with the baseline, BP Non-Preserving is more effective for improving both $R_\mathrm{hub}$ and $|$FVS$|$. 
Moreover, from the difference between the results in Non-Preserving and Preserving, it suggests that modification of the degree distribution significantly affects both $R_\mathrm{hub}$ and $|$FVS$|$.

\begin{figure}[h!]
  \centering
  \includegraphics[width=.7\linewidth]{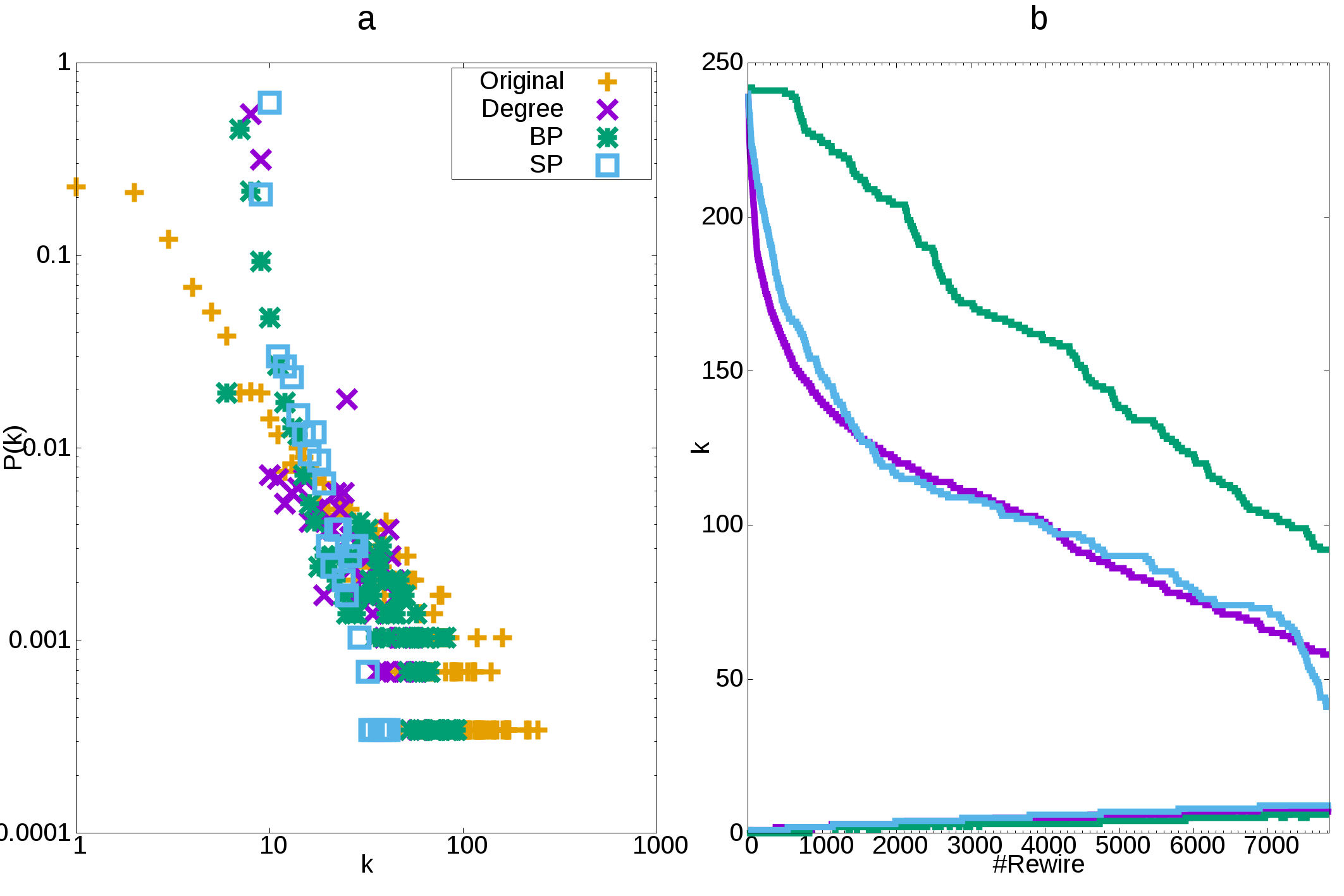}
  \caption{Change of the degree distributions by the rewirings in Non-Preserving.
      (a) Degree distributions in original and after rewiring networks, (b) Maximum and minimum degrees vs. the number of rewiring in Degree, BP, and SP Non-Preserving for OpenFlights.
      The above three lines show the maximum degrees. The below three lines show the minimum degrees. Violet, green, and light blue denote Degree, BP, and SP Non-Preserving. Orange denotes the original degree distribution.}
      \label{flog3}
\end{figure}

We find that the rewirings in Non-Preserving commonly make the network more homogeneous and reduce the fraction of the high degree nodes as follows.
Figure \ref{flog3}a shows the initial degree distribution for OpenFlights and the modified degree distribution after rewired 7800 times by each method.

The initial distribution has a long-tailed distribution characterized as a scale-free property, in which the minimum and maximum degrees are 1 and 242.
After rewiring by the three methods, a gap between the minimum and maximum degree becomes smaller.
In particular, SP Non-Preserving modifies to the very narrow gap with the minimum and maximum degree of 9 and 41 in which the 61\% occupy the nodes with degree 10. 
Figure \ref{flog3}b shows that the maximum degree is lower than 100.
The methods of Degree, BP, and SP Non-Preserving also decrease the gap between the minimum and maximum degrees.
Thus, it is suggest that reducing the gap of degrees leads to improve both $R_{\mathrm{hub}}$ and $|\mathrm{FVS}|$ significantly.

The increases in $R_{\mathrm{hub}}$ and $|\mathrm{FVS}|$ by BP Non-Preserving are partly due to changing the degree distributions as reducing large degree nodes.
On the other hand, the increases in them by BP Preserving are only due to enhancing loops with preserving degrees. 
However, the values of $R_{\mathrm{hub}}$ and $|\mathrm{FVS}|$ in BP Non-Preserving are larger than ones in BP Preserving.

\subsection*{Relation between the robustness and the size of FVS}
In more detailed comparisons, we discuss a relation between $R_\mathrm{hub}$ and $|$FVS$|$ in Preserving and Non-Preserving. 
At the left in Figs. \ref{flog2}ace for Preserving, we compare the ordering of BP, Degree, SP Preserving, and WuHolme by the maximum value of each index. 
It is $\mathrm{BP}>\mathrm{WuHolme}>\mathrm{Degree}>\mathrm{SP}$ for $R_\mathrm{hub}$, $\mathrm{BP}>\mathrm{Degree}>\mathrm{WuHolme}>\mathrm{SP}$ for $|$FVS$|$, and $\mathrm{Degree}>\mathrm{BP}>\mathrm{WuHolme}>\mathrm{SP}$ for $r$. 
The order for $R_\mathrm{hub}$ and $|$FVS$|$ are almost the same, only the order of Degree and WuHolme are exchanged. 
As shown in Table \ref{table1}, $R_\mathrm{hub}$ and $|$FVS$|$ have a very strong correlation coefficient of 0.970 in Preserving. 
On the other hand, the correlation coefficient between $R_\mathrm{hub}$ and $r$ is 0.762.
It is lower than that of $R_\mathrm{hub}$ and $|$FVS$|$.
These values suggest that $|$FVS$|$ is more strongly related to $R_\mathrm{hub}$ than $r$.

The difference in the ordering is more remarkable in Non-Preserving.
At the right in Figs. \ref{flog2}bdf for Non-Preserving, in BP, Degree, SP Non-Preserving and WuHolme, it is $\mathrm{Degree}>\mathrm{BP}>\mathrm{SP}>\mathrm{WuHolme}$ for both $R_\mathrm{hub}$ and $|$FVS$|$, while $\mathrm{Degree}>\mathrm{BP}>\mathrm{WuHolme}>\mathrm{SP}$ for $r$. 
Furthermore, as shown in Table \ref{table1}, the correlation coefficient in Non-Preserving is 0.980 for $R_\mathrm{hub}$ and $|$FVS$|$.
It is slightly larger than that in Preserving.
On the other hand, the correlation between $R_\mathrm{hub}$ and $r$ is 0.527, which is smaller than that in Preserving. 
From these results, the correlation between $R_\mathrm{hub}$ and $r$ in Non-Preserving becomes weaker than that in Preserving. 
Thus, $|$FVS$|$ is more strongly related to $R_\mathrm{hub}$ than $r$. 

\begin{table}[h!]
\centering
\caption{The correlation coefficient between $R_\mathrm{hub}$ and $r$, and $R_\mathrm{hub}$ and $|\mathrm{FVS}|$ after rewiring.}
    \begin{tabular}{l|rr}
    & $R_\mathrm{hub}$ and $r$ & $R_\mathrm{hub}$ and $|\mathrm{FVS}|$ \\ \hline
    Preserving & 0.762 & 0.970 \\
    Non-Preserving & 0.527 & 0.980 \\
    \end{tabular}
    \label{table1}
\end{table}

\subsection*{Strongly robust networks with negative degree-degree correlations}
As known in the onion-like structure, it has been considered that networks with the moderate degree-degree correlations tend to be more robust \cite{schneider2011mitigation,tanizawa2012robustness}.
However, from the obtained results, we find that networks with the negative degree-degree correlations are possible to be highly robust.
Figures \ref{flog2}bf show that SP Non-Preserving (denoted by the light blue line) decreases $r$, while increases $R_\mathrm{hub}$. 
SP Non-Preserving modifies it negatively up to -0.187, showing as the light blue line at the right in Fig. \ref{flog2}f. 
However, at the right in Fig. \ref{flog2}b, SP Non-Preserving increases $R_{\mathrm{hub}}$ at the almost same level as both Degree and BP Non-Preserving. 
Note that both Degree and BP Non-Preserving make the degree-degree correlations positive over the baseline.
These obtained results are commonly found for other networks (See Additional File).

\section{Summary}
This study proposes a strategy for enhancing loops in increasing the size of FVS to improve the robustness of connectivity. 
We consider two kinds of rewirings for enhancing loops as BP Preserving and BP Non-Preserving.
The rewiring in Preserving does not change each node's degree, while the rewiring in Non-Preserving changes the degree.
We obtain similar results in applying our proposed and conventional rewirings to several real networks (See the Additional Files).
From the results, BP Preserving increases the size of FVS effectively. 
It also improves the robustness index to the level as the same or more than the conventional best.
Thus, enhancing loops is a useful strategy for improving robustness. 
On the other hand, BP Non-Preserving increases the robustness index and the size of FVS much more than the conventional best in Preserving.
Moreover, the other rewirings in Non-Preserving also increase them as the same, and commonly reduce the gap of maximum and minimum degrees.
Therefore, it is suggested that reducing the difference in degrees strongly affects increasing the robustness index and the size of FVS.
Note that the results in BP Non-Preserving are partly due to changing the degree distributions, while ones in BP Preserving are only due to enhancing loops.

In addition, we discuss the relation between the robustness and the size of FVS. 
The size of FVS is more strongly related to the robustness than the degree-degree correlations in both Preserving and Non-Preserving.
We also find that existing of strongly robust networks with the negative degree-degree correlations is possible.
Therefore, we suggest that enhancing loops is more essential for improvements of robustness than the degree-degree correlations.

\section*{Acknowledgements}
  This research is supported in part by JSPS KAKENHI Grant Number JP.17H01729.

\bibliographystyle{bmc-mathphys}  
\bibliography{ms}
\end{document}


\maketitle

\appendix
\renewcommand{\thetable}{S\arabic{table}}
\renewcommand{\thefigure}{S\arabic{figure}}

\section*{Appendix}
For 10 real networks shown in Table \ref{datas_index}, we apply edge-rewiring methods: our proposed methods (BP), a method with smaller degrees instead of smaller $q^0_i$ of BP (Degree), SP \cite{chan2016optimizing}, and WuHolme \cite{wu2011onion}.
As a preprocessing, we transform the network data into undirected, unweighted, and no self-loop and no multiple edges, and extract the giant component.
In the Figures, we compare the effects on the robustness index $R_{\mathrm{hub}}$ \cite{schneider2011mitigation}, the approximate size of FVS by Ref \cite{zhou2013spin}, and the degree-degree correlations $r$ \cite{newman2002assortative} versus the number of rewiring.
In addition, to show the modifications in the degrees by rewiring in Non-Preserving, we show the gap between the maximum and minimum degrees versus the number of rewiring in original and after rewiring networks.

\begin{figure}
{\LARGE Power Grid}
\centering
\includegraphics[width=\textwidth]{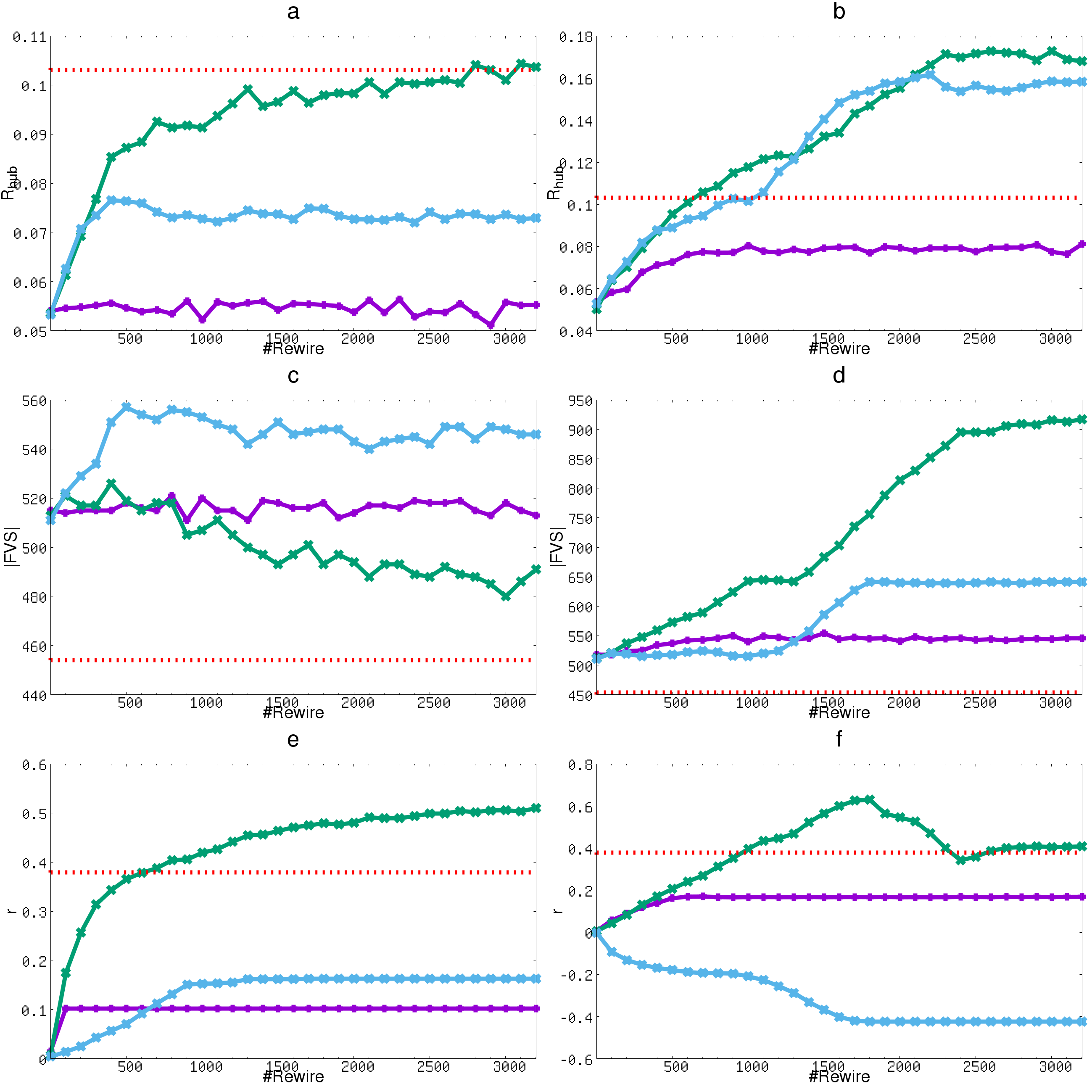}
\caption{Power Grid \cite{watts1998collective}. Comparison of the robustness index $R_{\mathrm{hub}}$, the size approximate of FVS, and the degree-degree correlation coefficient $r$ vs. the number of rewiring. (Left: a, c, e) Rewirings in Preserving. Violet, green, and light blue solid lines denote the result by Degree, BP, and SP Preserving, respectively. The red dot line indicates a baseline of the conventional best. (Right: b, d, f) Rewirings in Non-Preserving. Violet, green, and light blue solid lines denote the result by Degree, BP, and SP Non-Preserving, respectively.}
\label{s3-1}
\end{figure}

The robustness and the size of FVS are more strongly related to each other than the degree-degree correlations. 
There is an exception for Power Grid shown in Figs. \ref{s3-1}ac. 
BP Preserving (denoted by the green line at the left) increases the robustness over the baseline but decreases the size of FVS. 
We consider that it is caused by a special property of Power Grid with a smaller average degree and maximum degree, but a larger diameter shown in Table \ref{datas_index}.

\begin{table}[hp]
\centering
    \begin{tabular}{l|rr|r|rrr|r|r|r|r}
    Network & $N$ & $M$ & $r$ & Min $k$ & $<k>$ & Max $k$ & $D$ & Figs. & Refs. & URL \\ \hline
    AirTraffic & 1226 & 2408 & -0.015 & 1 & 3.9 & 34 & 17 & \ref{s1-1}, \ref{s1-2} & \cite{kunegis2013konect}& \href{http://konect.uni-koblenz.de/networks/maayan-faa}{url} \\
    E-mail & 1133 & 5451 & 0.078 & 1 & 9.6 & 71 & 8 & \ref{s2-1}, \ref{s2-2} &\cite{guimera2003self}& \href{http://deim.urv.cat/~alexandre.arenas/data/welcome.htm}{url}\\
    PowerGrid & 4941 & 6594 & 0.003 & 1 & 2.7 & 19 & 46 & \ref{s3-1}, \ref{s3-2} & \cite{watts1998collective}& \href{http://www-personal.umich.edu/~mejn/netdata/}{url} \\
    Yeast & 2224 & 6609 & -0.105 & 1 & 5.9 & 64 & 11 & \ref{s4-1}, \ref{s4-2} & \cite{bu2003topological}& \href{http://www.weizmann.ac.il/mcb/UriAlon/download/collection-complex-networks}{url}\\
    Japanese & 2698 & 7995 & -0.259 & 1 & 5.9 & 725 & 8 & \ref{s5-1}, \ref{s5-2} &  \cite{milo2004superfamilies} & \href{http://www.weizmann.ac.il/mcb/UriAlon/download/collection-complex-networks}{url}\\
    Hamster & 1788 & 12476 & -0.089 & 1 & 14.0 & 272 & 14 & \ref{s6-1}, \ref{s6-2} &  \cite{kunegis2013konect} & \href{http://konect.uni-koblenz.de/networks/petster-friendships-hamster}{url}\\
    GRQC & 4158 & 13422 & 0.639 & 1 & 6.5 & 81 & 17 & \ref{s7-1}, \ref{s7-2} &  \cite{leskovec2007graph} & \href{http://snap.stanford.edu/data/ca-GrQc.html}{url}\\
    UCIrvine & 1893 & 13835 & -0.188 & 1 & 14.6 & 255 & 8 & \ref{s8-1}, \ref{s8-2} & \cite{kunegis2013konect, opsahl2009clustering} & \href{http://konect.uni-koblenz.de/networks/opsahl-ucsocial}{url}\\
    OpenFlights & 2905 & 15645 & 0.049 & 1 & 10.8 & 242 & 14 & \ref{s9-1}, \ref{s9-2} & \cite{kunegis2013konect,opsahl2010node} & \href{http://konect.uni-koblenz.de/networks/opsahl-openflights}{url}\\
    Polblogs & 1222 & 16714 & -0.221 & 1 & 27.4 & 351 & 8 & \ref{s10-1}, \ref{s10-2} & \cite{adamic2005political} & \href{http://www-personal.umich.edu/~mejn/netdata/}{url}\\
    \end{tabular}
\caption{Basic properties for real networks after preprocessing. From the left, we note the name of the network, the number of nodes, the number of edges, the degree-degree correlations, the minimum degree, the average degree, the maximum degree, the diameter, figures, references, and available URL to download the data. }
\label{datas_index}
\end{table}
\newpage

\begin{figure}[h]
{\LARGE AirTraffic}
\centering
\includegraphics[width=\textwidth]{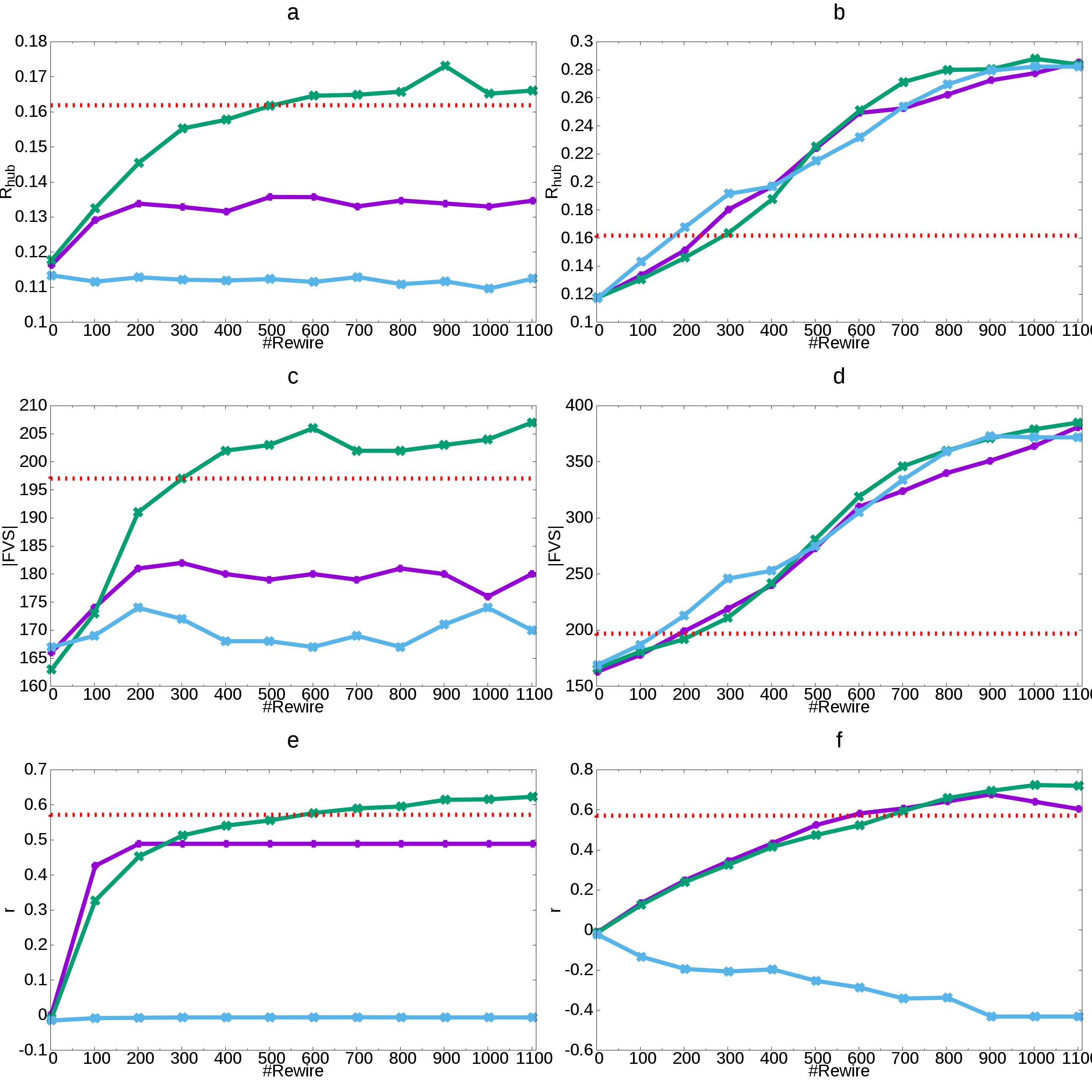}
\caption{AirTraffic \cite{kunegis2013konect}. Comparison of the robustness index $R_{\mathrm{hub}}$, the size approximate of FVS, and the degree-degree correlation coefficient $r$ vs. the number of rewiring. (Left: a, c, e) Rewirings in Preserving. Violet, green, and light blue solid lines denote the result by Degree, BP, and SP Preserving, respectively. The red dot line indicates a baseline of the conventional best. (Right: b, d, f) Rewirings in Non-Preserving. Violet, green, and light blue solid lines denote the result by Degree, BP, and SP Non-Preserving, respectively.}
\label{s1-1}
\end{figure}

\begin{figure}[h]
{\LARGE AirTraffic}
\centering
\includegraphics[width=\textwidth]{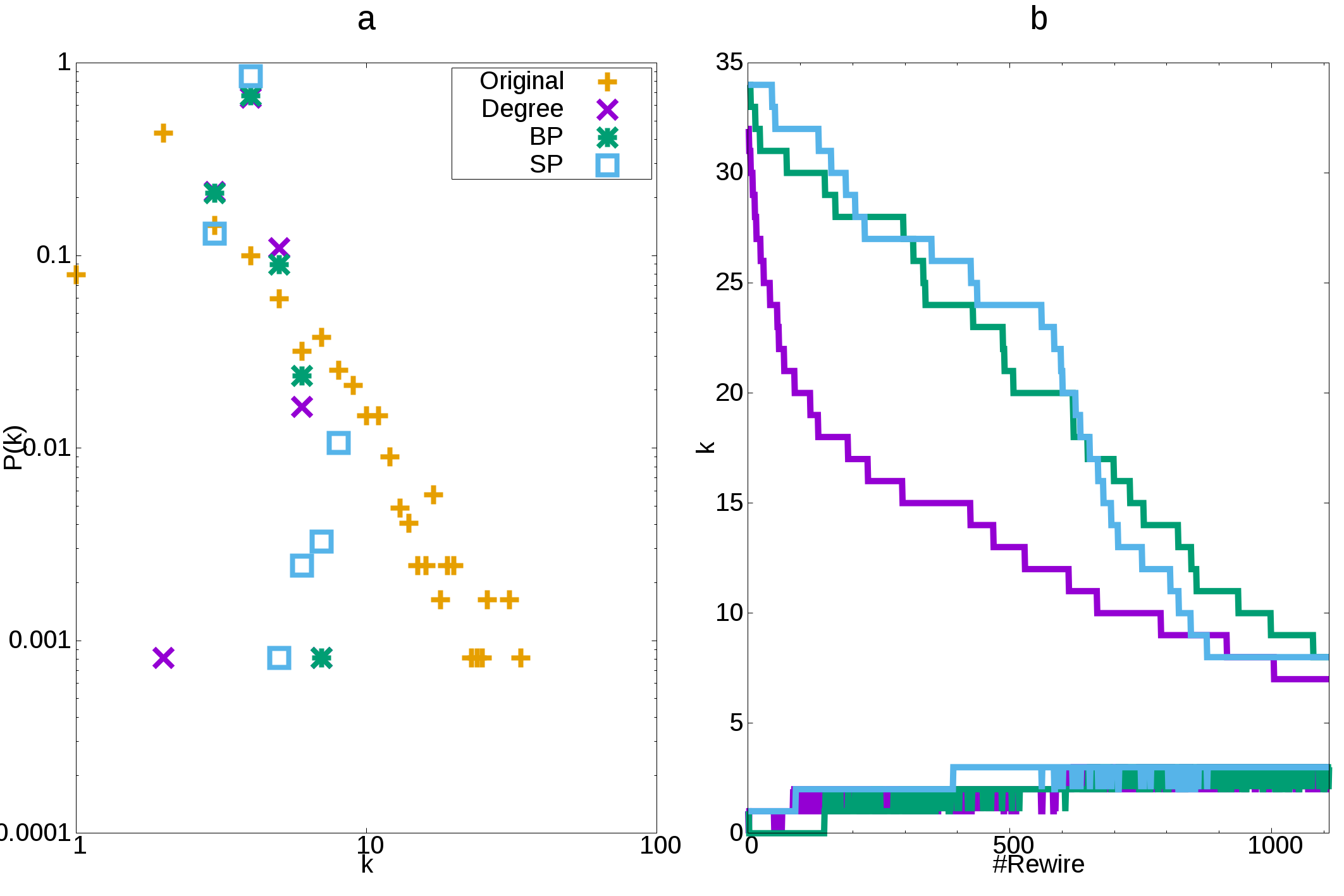}
\caption{AirTraffic \cite{kunegis2013konect}. (a) Degree distributions in original and after rewiring networks, (b) Maximum and minimum degrees vs. the number of rewiring in Degree, BP, and SP Non-Preserving. The above three lines show the maximum degrees. The below three lines show the minimum degrees. Violet, green, and light blue denote Degree, BP, SP Non-Preserving. Orange denotes the original degree distribution.}
\label{s1-2}
\end{figure}

\begin{figure}
{\LARGE E-mail}
\centering
\includegraphics[width=\textwidth]{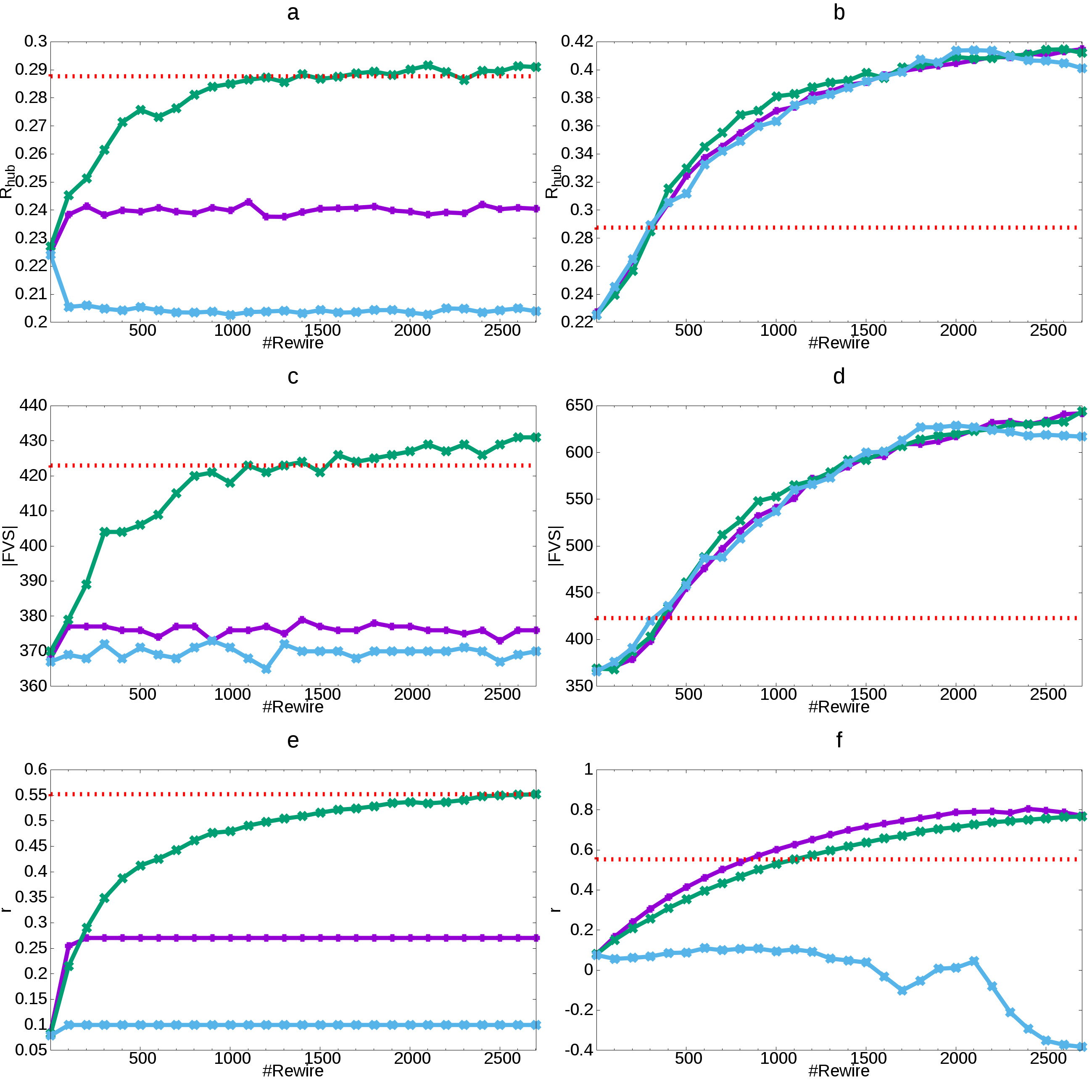}
\caption{E-mail \cite{guimera2003self}. Comparison of the robustness index $R_{\mathrm{hub}}$, the size approximate of FVS, and the degree-degree correlation coefficient $r$ vs. the number of rewiring. (Left: a, c, e) Rewirings in Preserving. Violet, green, and light blue solid lines denote the result by Degree, BP, and SP Preserving, respectively. The red dot line indicates a baseline of the conventional best. (Right: b, d, f) Rewirings in Non-Preserving. Violet, green, and light blue solid lines denote the result by Degree, BP, and SP Non-Preserving, respectively.}
\label{s2-1}
\end{figure}

\begin{figure}
{\LARGE E-mail}
\centering
\includegraphics[width=\textwidth]{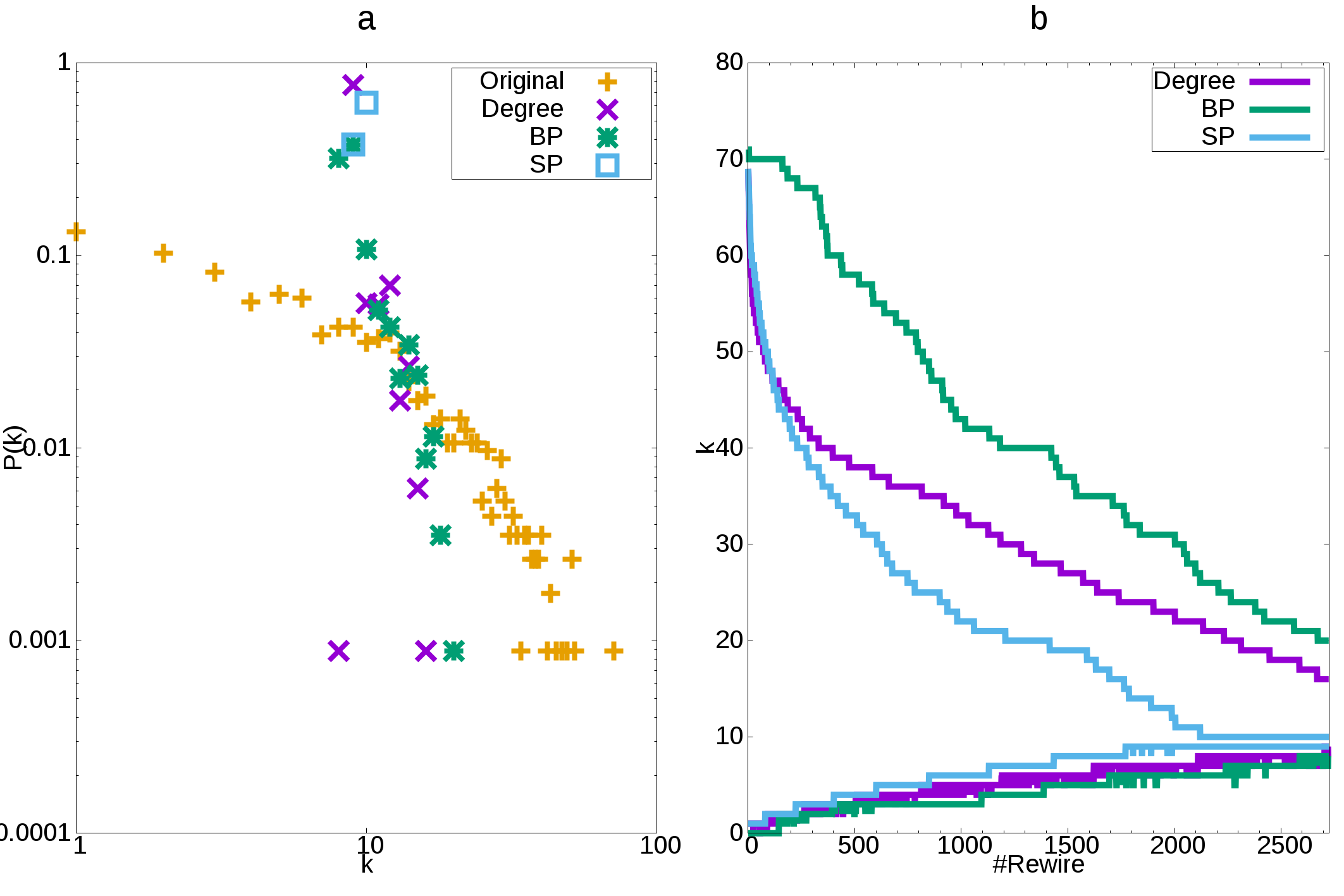}
\caption{E-mail \cite{guimera2003self}. (a) Degree distributions in original and after rewiring networks, (b) Maximum and minimum degrees vs. the number of rewiring in Degree, BP, and SP Non-Preserving. The above three lines show the maximum degrees. The below three lines show the minimum degrees. Violet, green, and light blue denote Degree, BP, SP Non-Preserving. Orange denotes the original degree distribution.}
\label{s2-2}
\end{figure}

\begin{figure}
{\LARGE Power Grid}
\centering
\includegraphics[width=\textwidth]{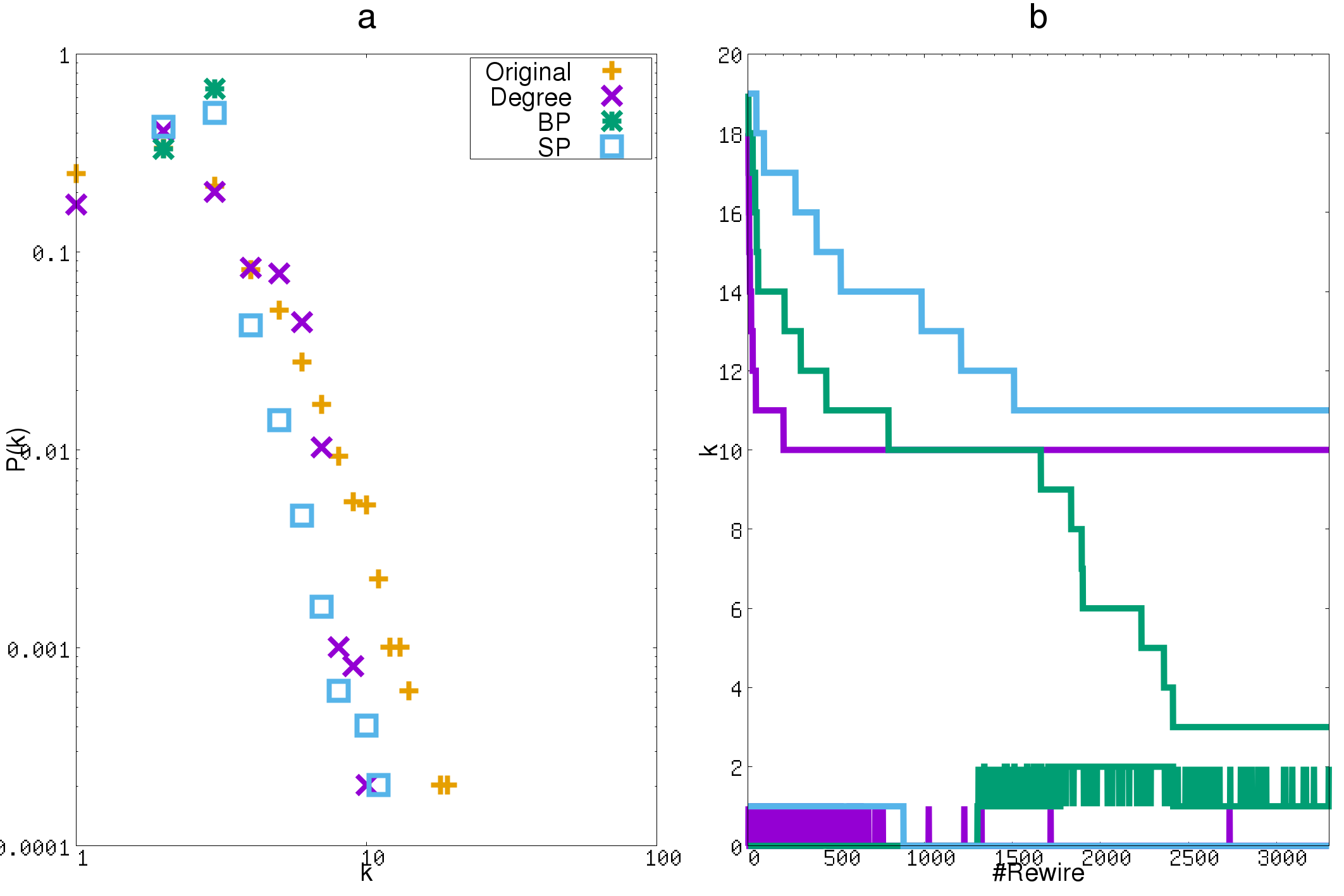}
\caption{Power Grid \cite{watts1998collective}. (a) Degree distributions in original and after rewiring networks, (b) Maximum and minimum degrees vs. the number of rewiring in Degree, BP, and SP Non-Preserving. The above three lines show the maximum degrees. The below three lines show the minimum degrees. Violet, green, and light blue denote Degree, BP, SP Non-Preserving. Orange denotes the original degree distribution.}
\label{s3-2}
\end{figure}

\begin{figure}
{\LARGE Yeast}
\centering
\includegraphics[width=\textwidth]{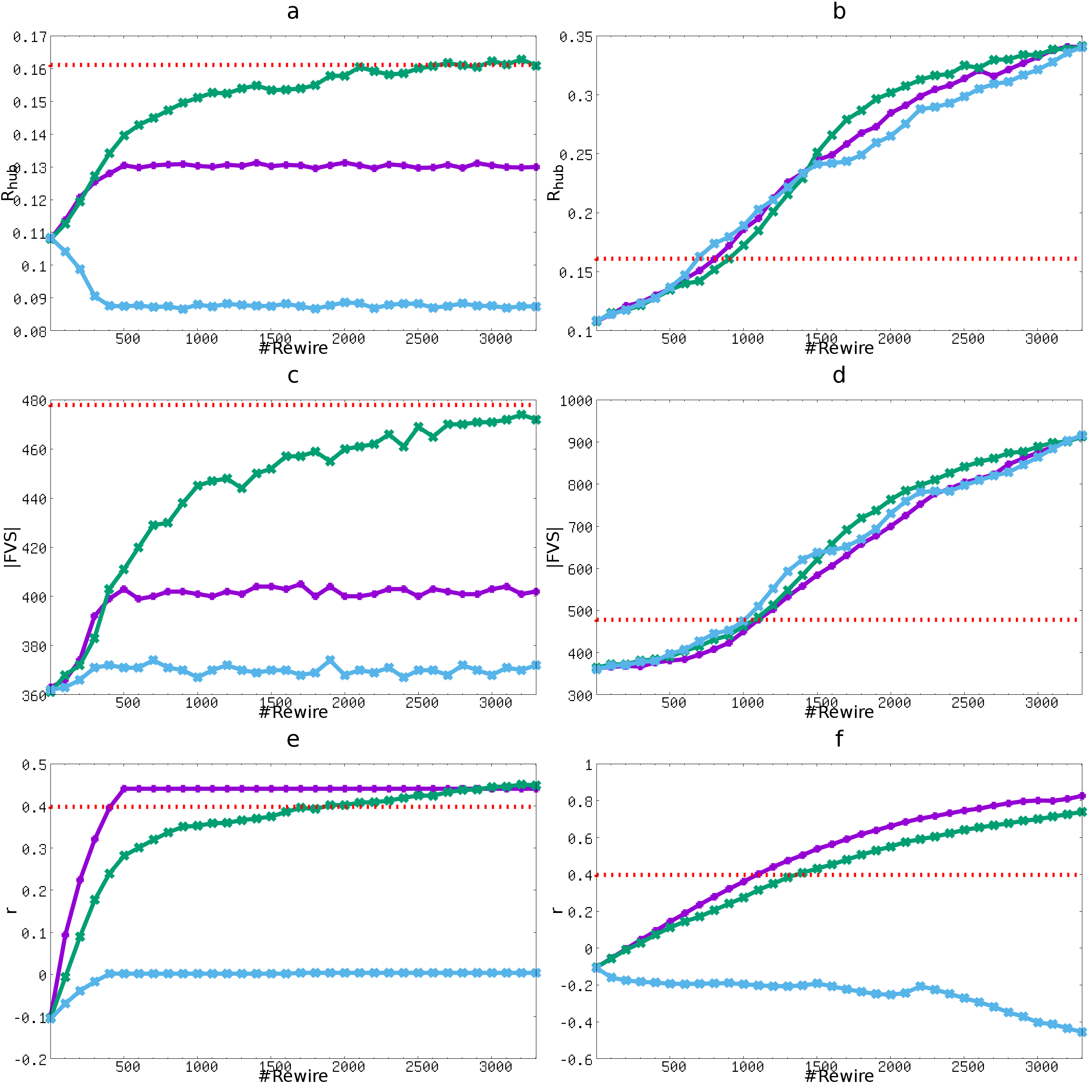}
\caption{Yeast \cite{bu2003topological}. Comparison of the robustness index $R_{\mathrm{hub}}$, the size approximate of FVS, and the degree-degree correlation coefficient $r$ vs. the number of rewiring. (Left: a, c, e) Rewirings in Preserving. Violet, green, and light blue solid lines denote the result by Degree, BP, and SP Preserving, respectively. The red dot line indicates a baseline of the conventional best. (Right: b, d, f) Rewirings in Non-Preserving. Violet, green, and light blue solid lines denote the result by Degree, BP, and SP Non-Preserving, respectively.}
\label{s4-1}
\end{figure}

\begin{figure}
{\LARGE Yeast}
\centering
\includegraphics[width=\textwidth]{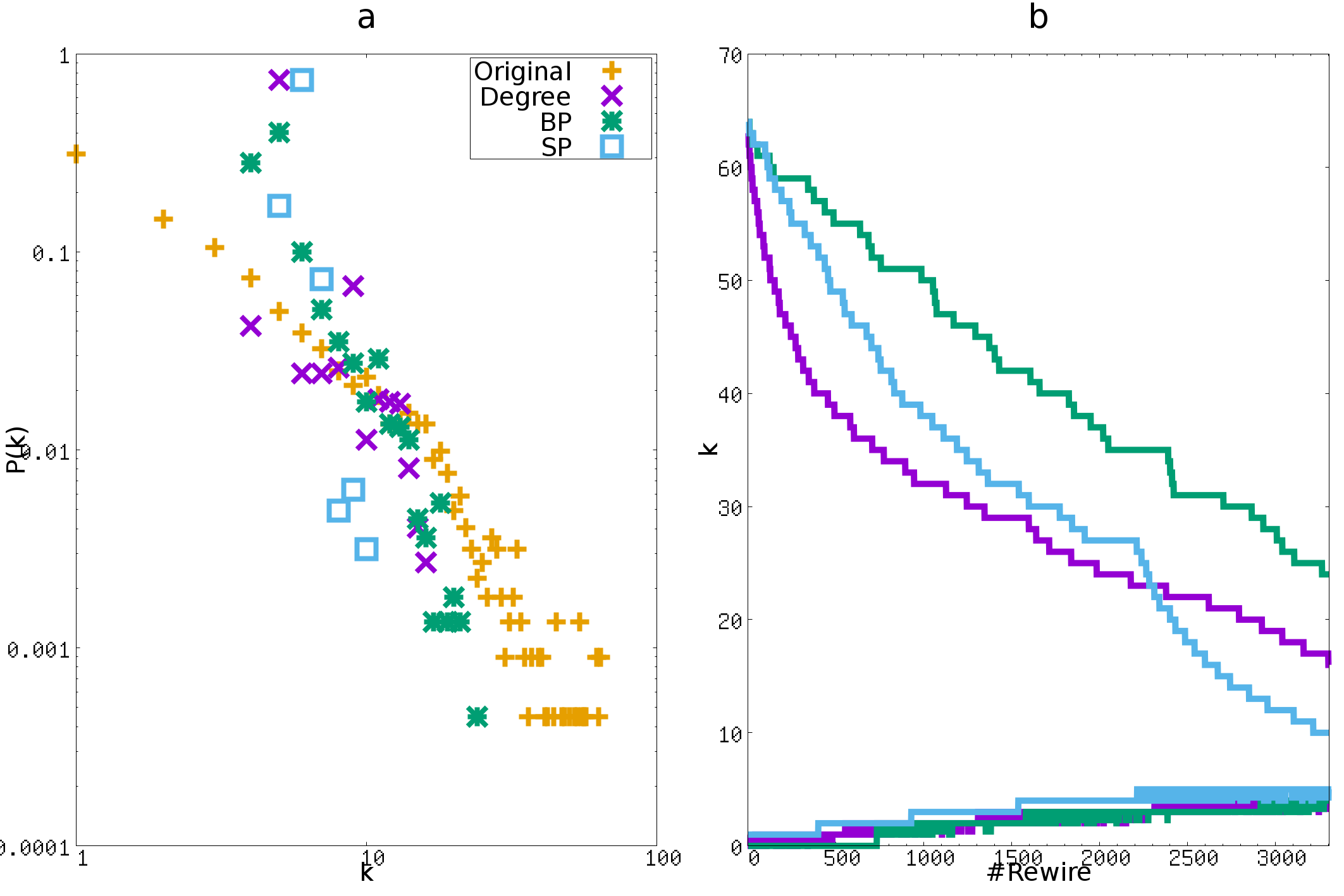}
\caption{Yeast \cite{bu2003topological}. (a) Degree distributions in original and after rewiring networks, (b) Maximum and minimum degrees vs. the number of rewiring in Degree, BP, and SP Non-Preserving. The above three lines show the maximum degrees. The below three lines show the minimum degrees. Violet, green, and light blue denote Degree, BP, SP Non-Preserving. Orange denotes the original degree distribution.}
\label{s4-2}
\end{figure}

\begin{figure}
{\LARGE Japanese}
\centering
\includegraphics[width=\textwidth]{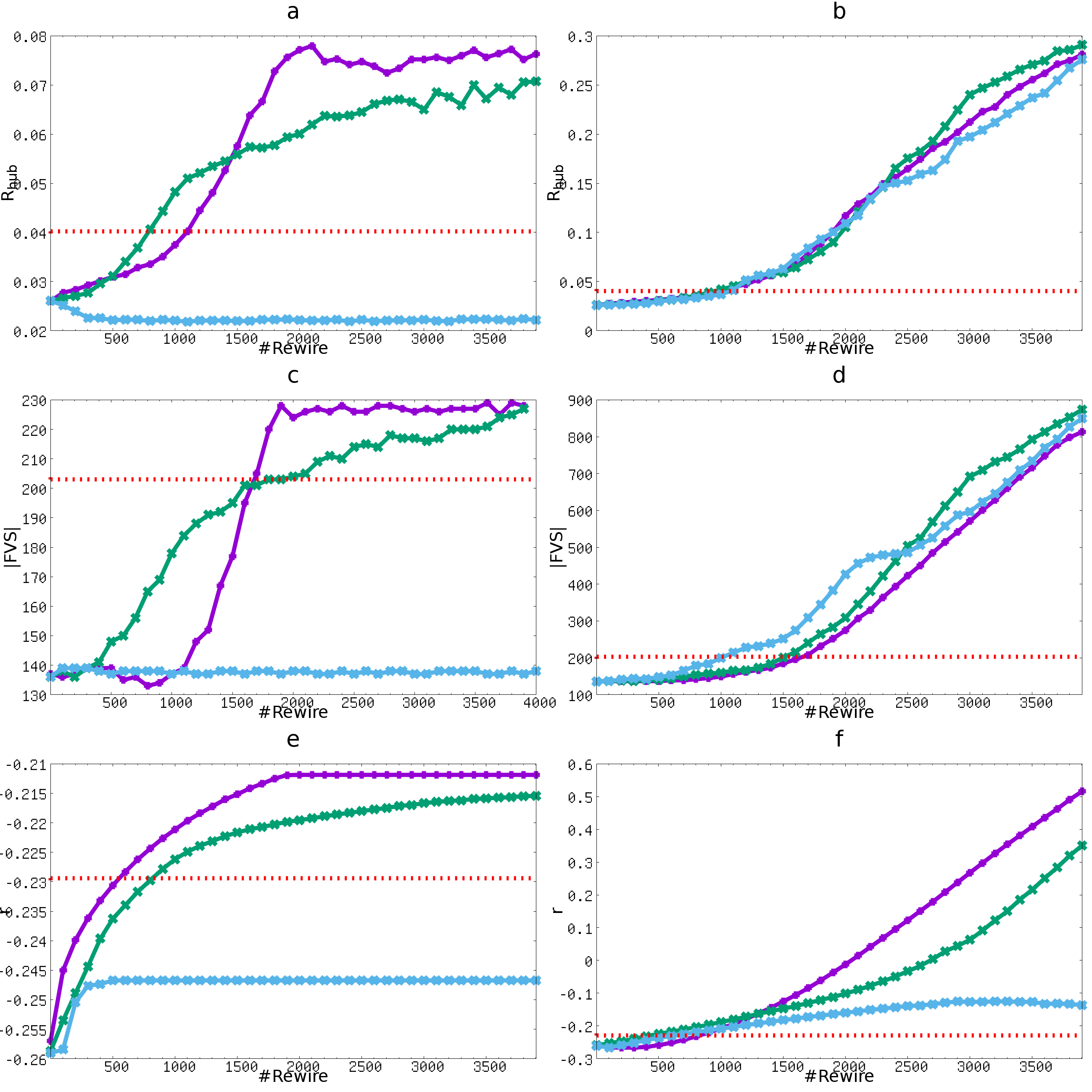}
\caption{Japanese \cite{milo2004superfamilies}. Comparison of the robustness index $R_{\mathrm{hub}}$, the size approximate of FVS, and the degree-degree correlation coefficient $r$ vs. the number of rewiring. (Left: a, c, e) Rewirings in Preserving. Violet, green, and light blue solid lines denote the result by Degree, BP, and SP Preserving, respectively. The red dot line indicates a baseline of the conventional best. (Right: b, d, f) Rewirings in Non-Preserving. Violet, green, and light blue solid lines denote the result by Degree, BP, and SP Non-Preserving, respectively.}
\label{s5-1}
\end{figure}

\begin{figure}
{\LARGE Japanese}
\centering
\includegraphics[width=\textwidth]{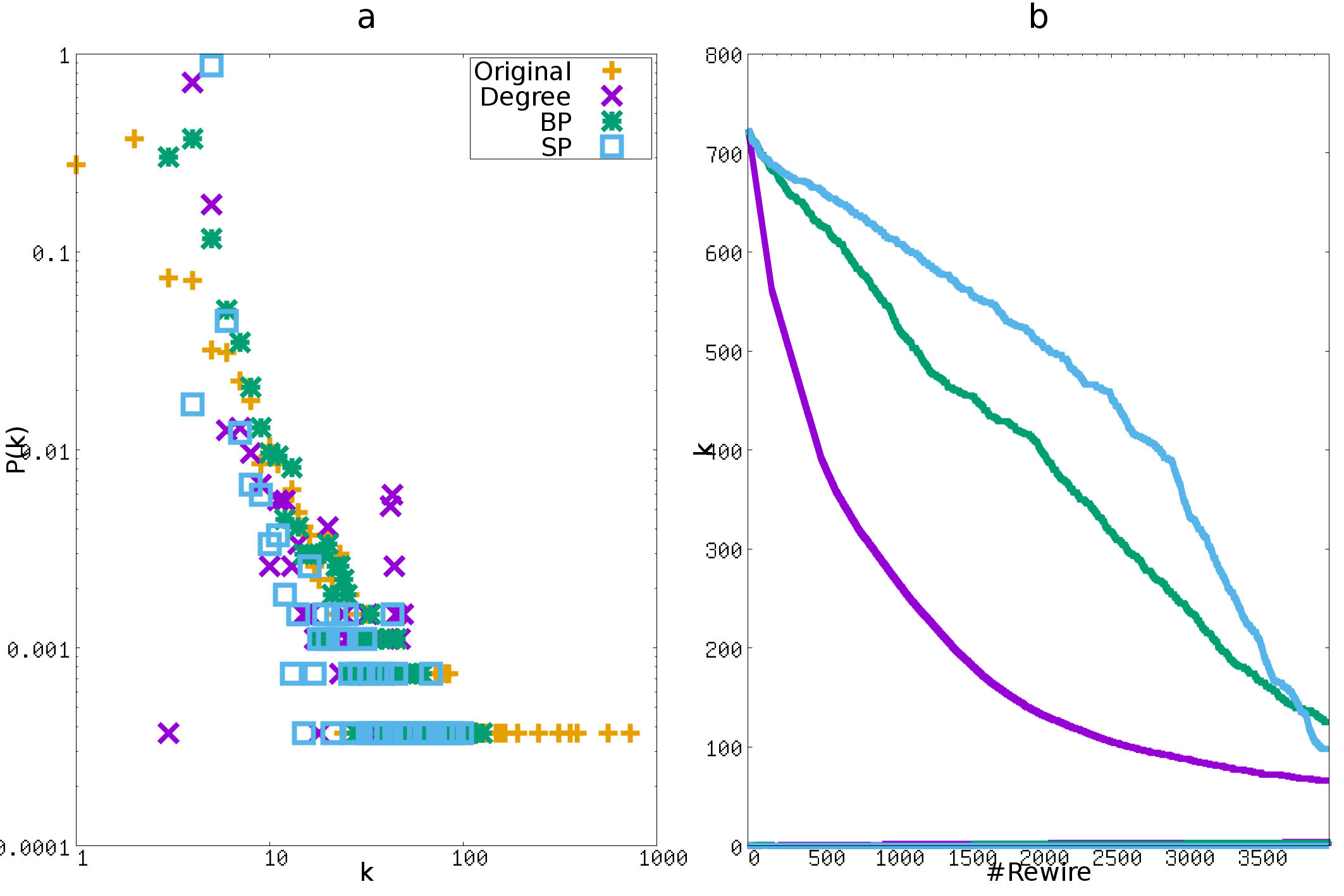}
\caption{Japanese \cite{milo2004superfamilies}. (a) Degree distributions in original and after rewiring networks, (b) Maximum and minimum degrees vs. the number of rewiring in Degree, BP, and SP Non-Preserving. The above three lines show the maximum degrees. The below three lines show the minimum degrees. Violet, green, and light blue denote Degree, BP, SP Non-Preserving. Orange denotes the original degree distribution.}
\label{s5-2}
\end{figure}

\begin{figure}
{\LARGE Hamster}
\centering
\includegraphics[width=\textwidth]{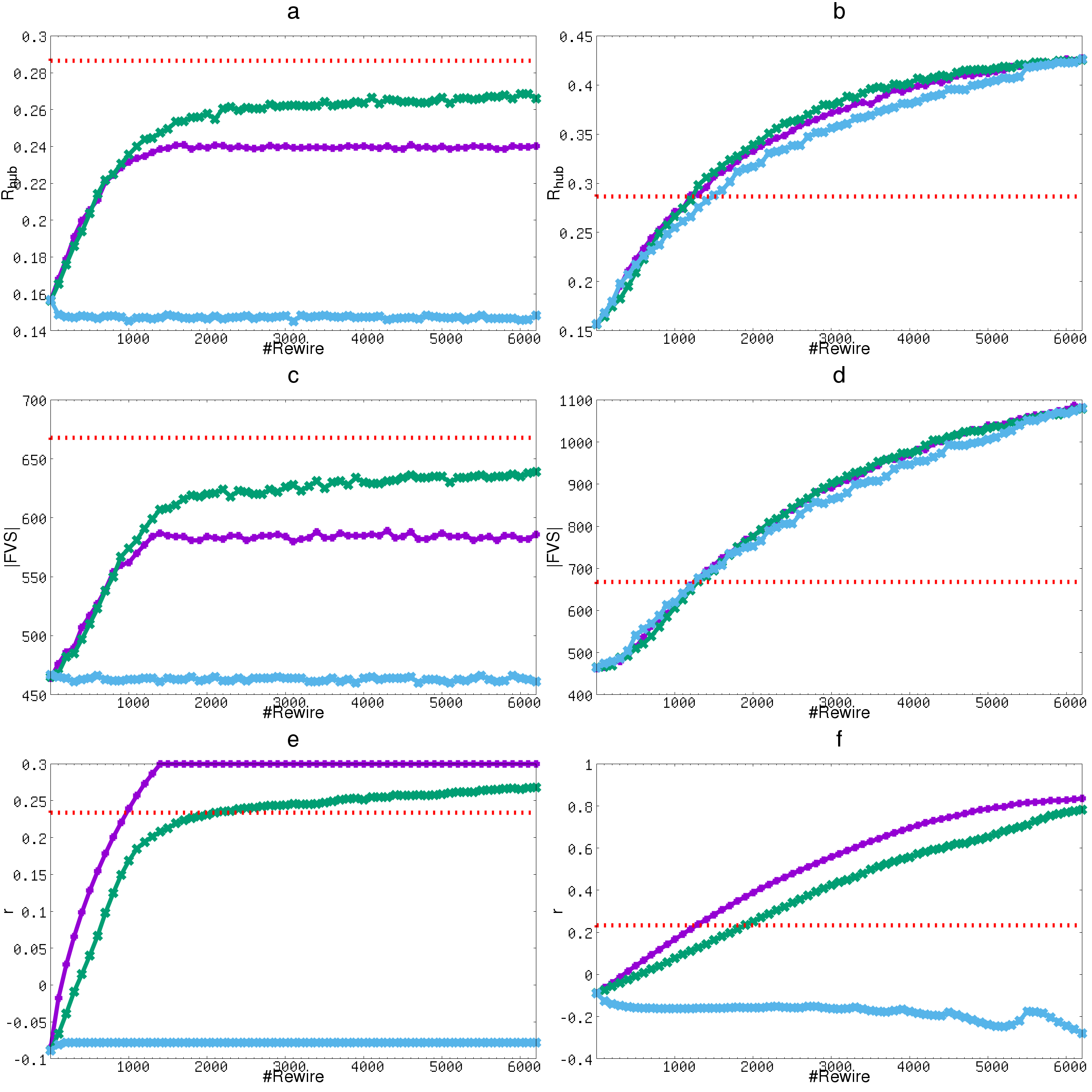}
\caption{Hamster \cite{kunegis2013konect}. Comparison of the robustness index $R_{\mathrm{hub}}$, the size approximate of FVS, and the degree-degree correlation coefficient $r$ vs. the number of rewiring. (Left: a, c, e) Rewirings in Preserving. Violet, green, and light blue solid lines denote the result by Degree, BP, and SP Preserving, respectively. The red dot line indicates a baseline of the conventional best. (Right: b, d, f) Rewirings in Non-Preserving. Violet, green, and light blue solid lines denote the result by Degree, BP, and SP Non-Preserving, respectively.}
\label{s6-1}
\end{figure}

\begin{figure}
{\LARGE Hamster}
\centering
\includegraphics[width=\textwidth]{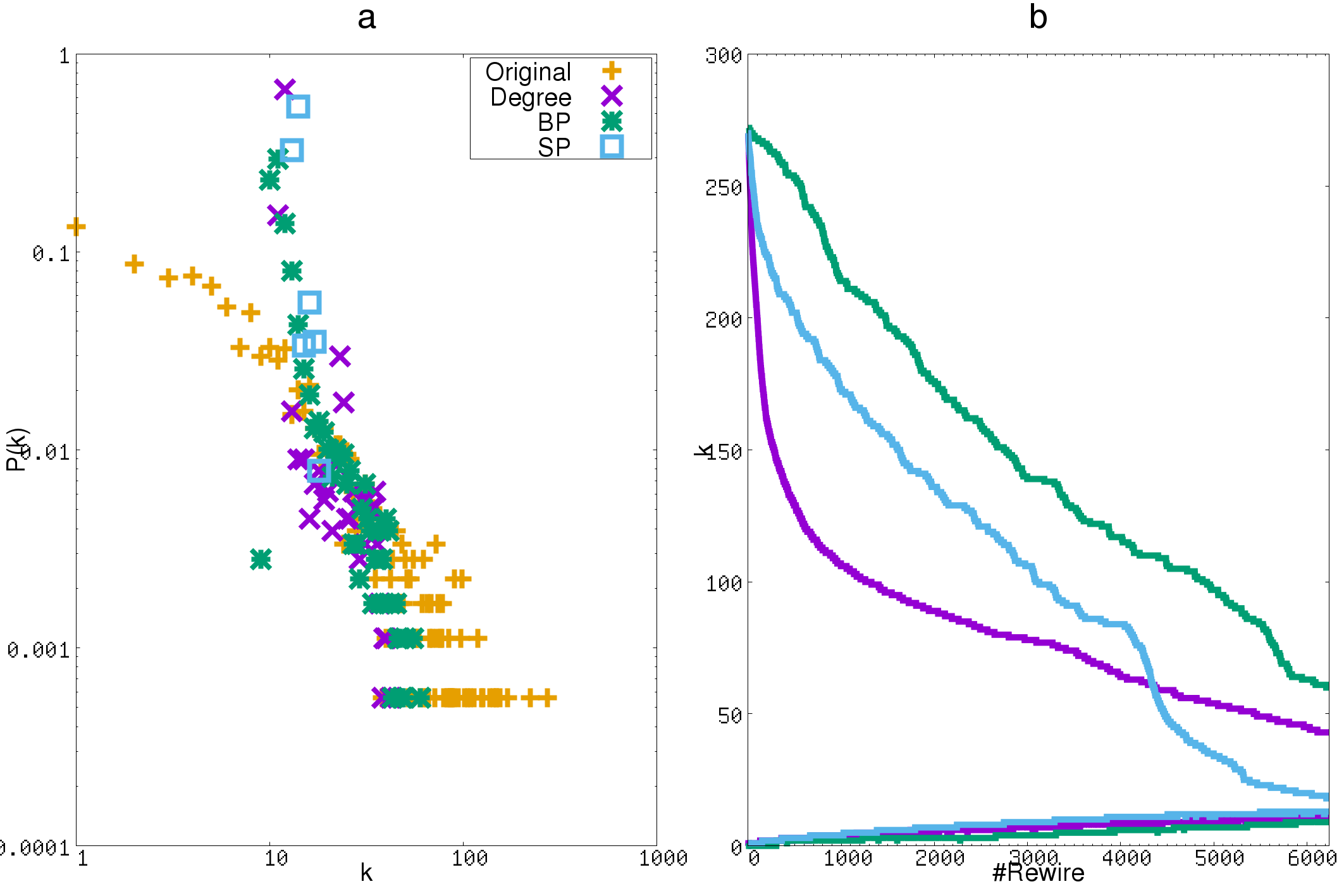}
\caption{Hamster \cite{kunegis2013konect}. (a) Degree distributions in original and after rewiring networks, (b) Maximum and minimum degrees vs. the number of rewiring in Degree, BP, and SP Non-Preserving. The above three lines show the maximum degrees. The below three lines show the minimum degrees. Violet, green, and light blue denote Degree, BP, SP Non-Preserving. Orange denotes the original degree distribution.}
\label{s6-2}
\end{figure}

\begin{figure}
{\LARGE GRQC}
\centering
\includegraphics[width=\textwidth]{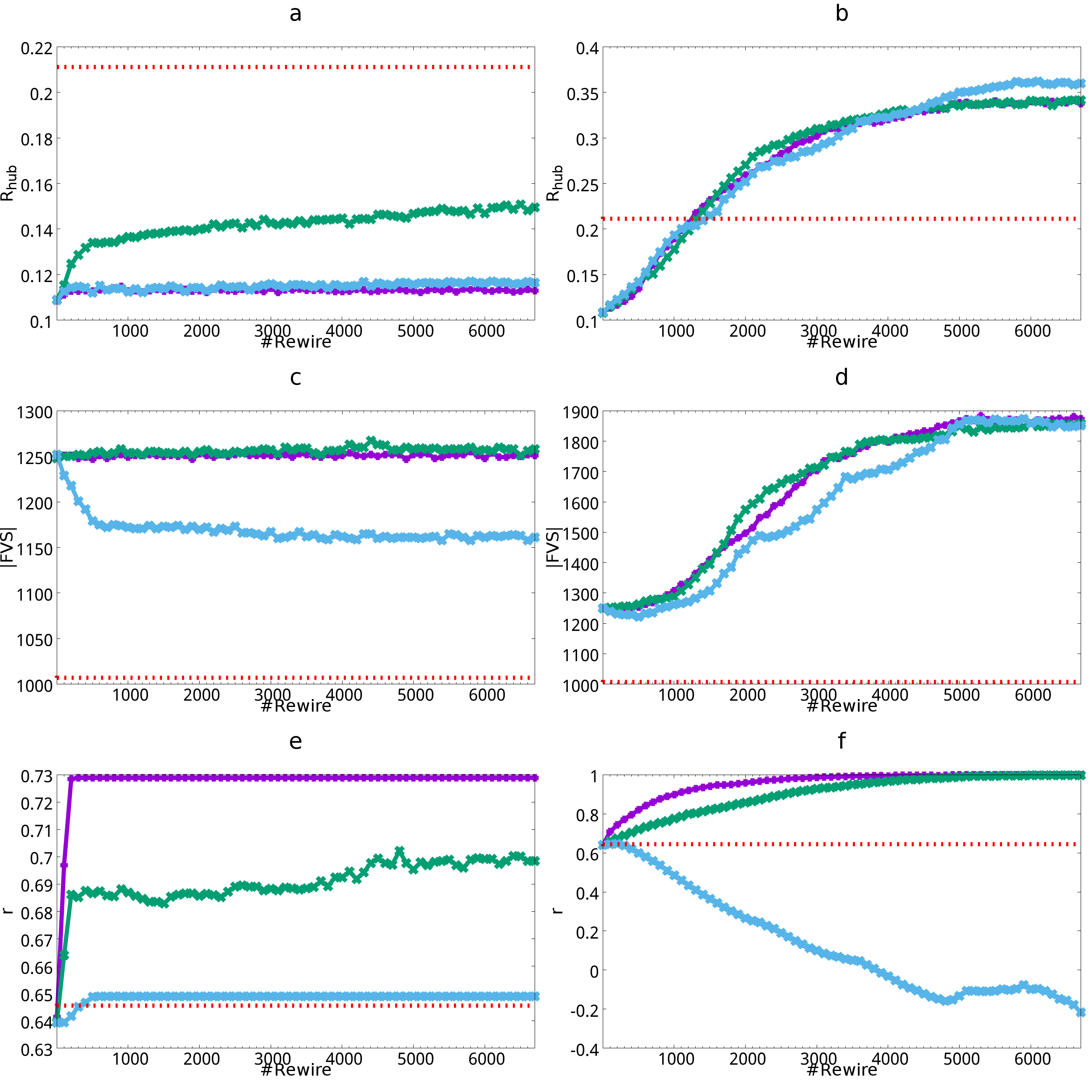}
\caption{GRQC \cite{leskovec2007graph}. Comparison of the robustness index $R_{\mathrm{hub}}$, the size approximate of FVS, and the degree-degree correlation coefficient $r$ vs. the number of rewiring. (Left: a, c, e) Rewirings in Preserving. Violet, green, and light blue solid lines denote the result by Degree, BP, and SP Preserving, respectively. The red dot line indicates a baseline of the conventional best. (Right: b, d, f) Rewirings in Non-Preserving. Violet, green, and light blue solid lines denote the result by Degree, BP, and SP Non-Preserving, respectively.}
\label{s7-1}
\end{figure}

\begin{figure}
{\LARGE GRQC}
\centering
\includegraphics[width=\textwidth]{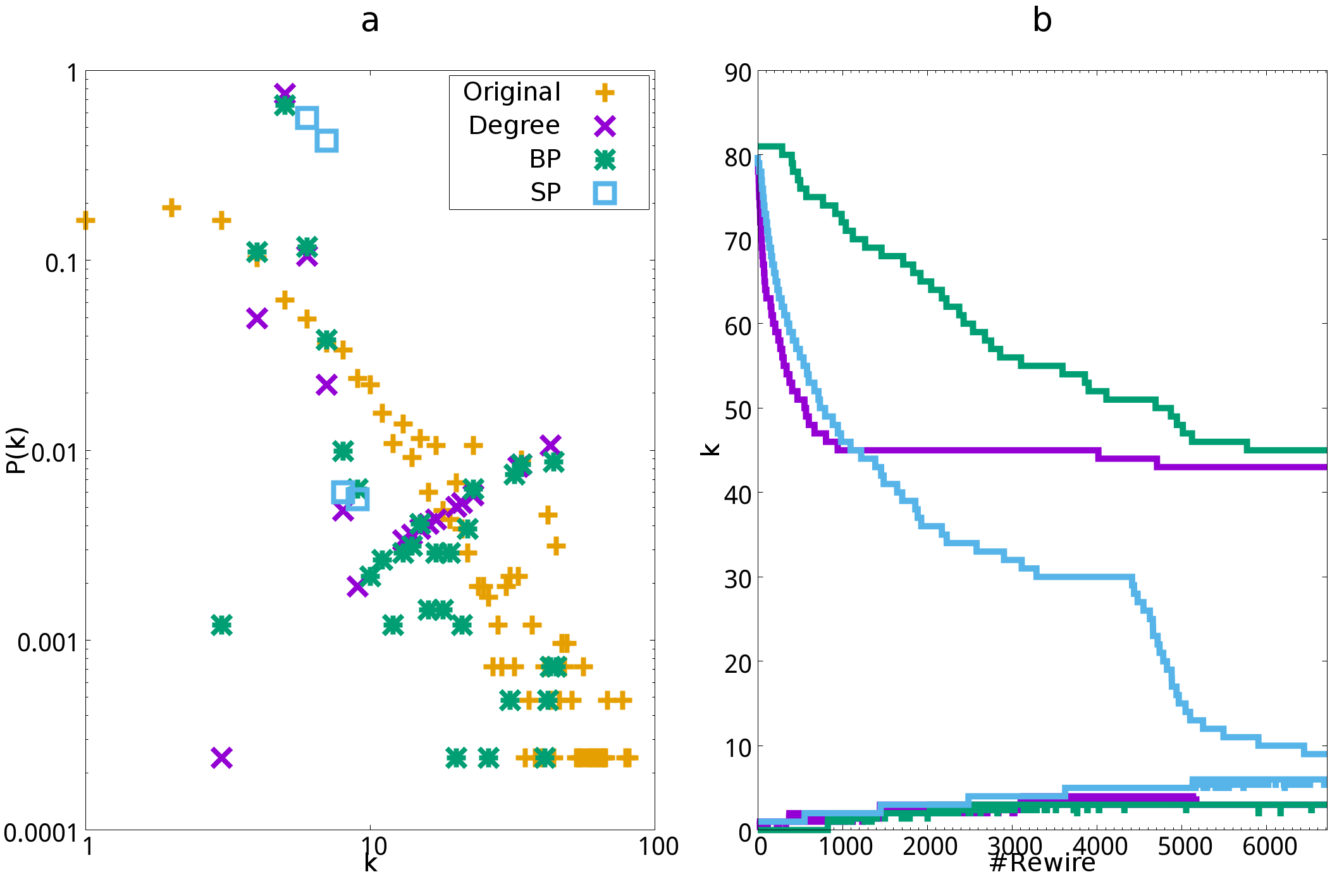}
\caption{GRQC \cite{leskovec2007graph}. (a) Degree distributions in original and after rewiring networks, (b) Maximum and minimum degrees vs. the number of rewiring in Degree, BP, and SP Non-Preserving. The above three lines show the maximum degrees. The below three lines show the minimum degrees. Violet, green, and light blue denote Degree, BP, SP Non-Preserving. Orange denotes the original degree distribution.}
\label{s7-2}
\end{figure}

\begin{figure}
{\LARGE UCIrvine}
\centering
\includegraphics[width=\textwidth]{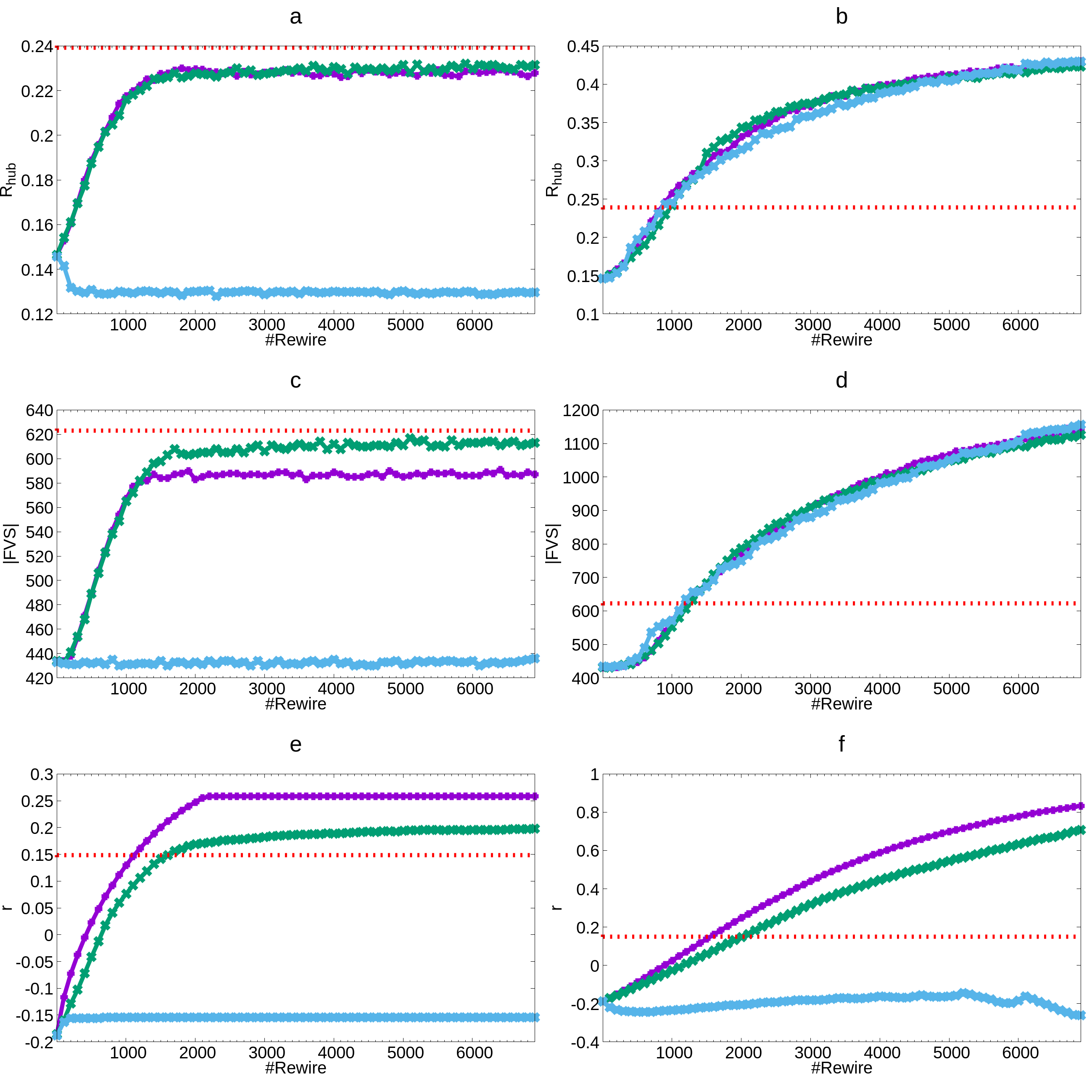}
\caption{UCIrvine \cite{kunegis2013konect,opsahl2009clustering}. Comparison of the robustness index $R_{\mathrm{hub}}$, the size approximate of FVS, and the degree-degree correlation coefficient $r$ vs. the number of rewiring. (Left: a, c, e) Rewirings in Preserving. Violet, green, and light blue solid lines denote the result by Degree, BP, and SP Preserving, respectively. The red dot line indicates a baseline of the conventional best. (Right: b, d, f) Rewirings in Non-Preserving. Violet, green, and light blue solid lines denote the result by Degree, BP, and SP Non-Preserving, respectively.}
\label{s8-1}
\end{figure}

\begin{figure}
{\LARGE UCIrvine}
\centering
\includegraphics[width=\textwidth]{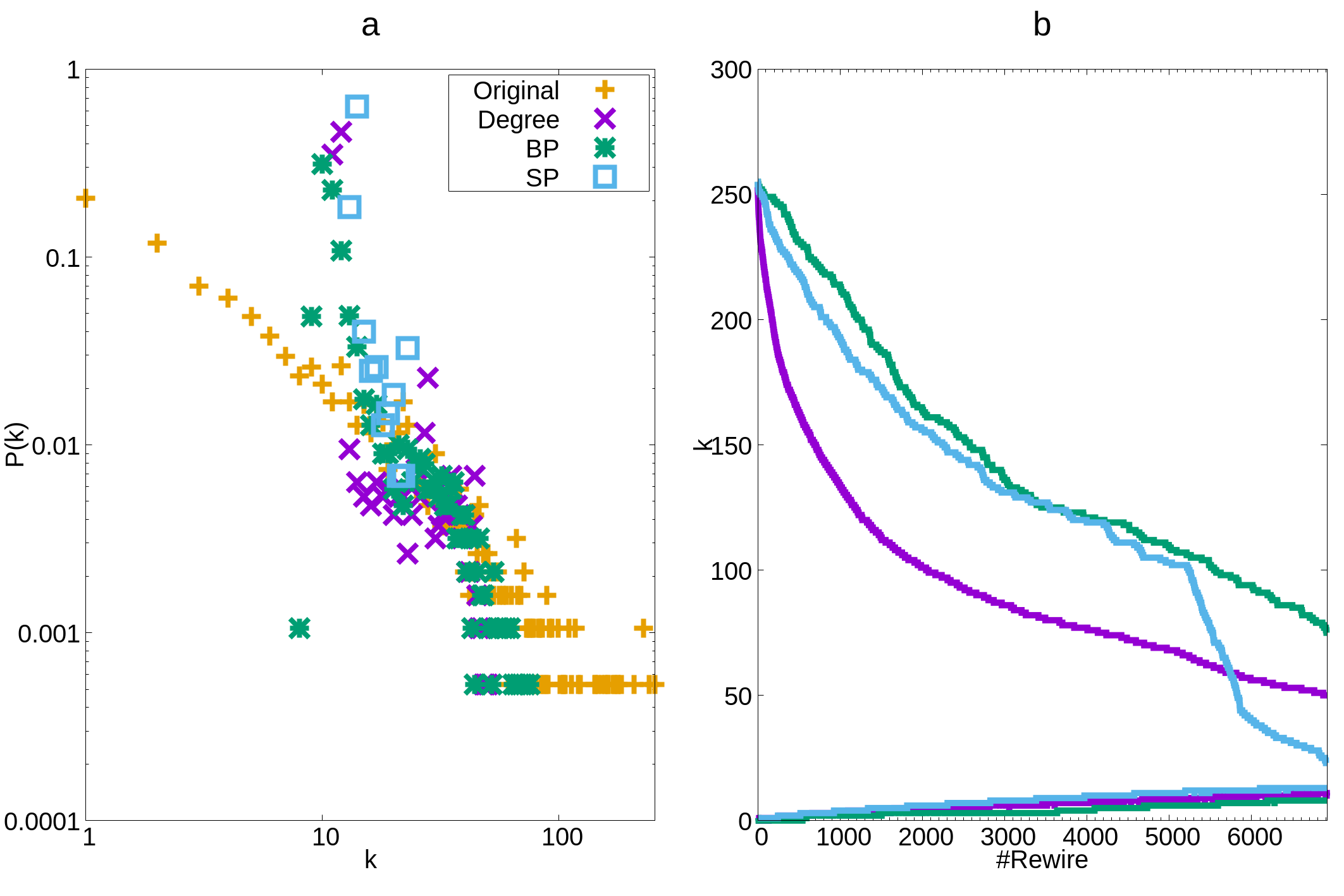}
\caption{UCIrvine \cite{kunegis2013konect,opsahl2009clustering}. (a) Degree distributions in original and after rewiring networks, (b) Maximum and minimum degrees vs. the number of rewiring in Degree, BP, and SP Non-Preserving. The above three lines show the maximum degrees. The below three lines show the minimum degrees. Violet, green, and light blue denote Degree, BP, SP Non-Preserving. Orange denotes the original degree distribution.}
\label{s8-2}
\end{figure}

\begin{figure}
{\LARGE OpenFlights}
\centering
\includegraphics[width=\textwidth]{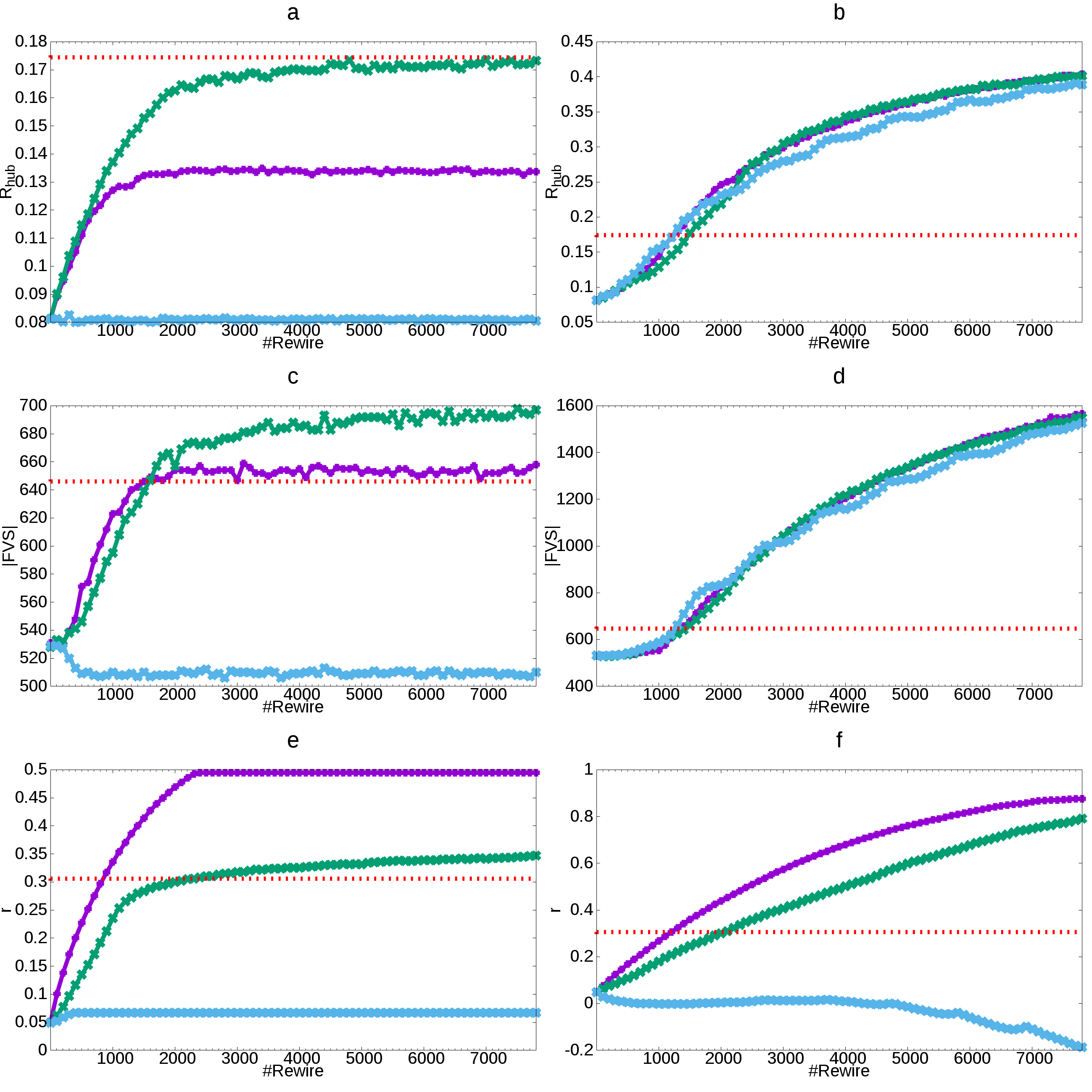}
\caption{OpenFlights \cite{kunegis2013konect,opsahl2010node}. Comparison of the robustness index $R_{\mathrm{hub}}$, the size approximate of FVS, and the degree-degree correlation coefficient $r$ vs. the number of rewiring. (Left: a, c, e) Rewirings in Preserving. Violet, green, and light blue solid lines denote the result by Degree, BP, and SP Preserving, respectively. The red dot line indicates a baseline of the conventional best. (Right: b, d, f) Rewirings in Non-Preserving. Violet, green, and light blue solid lines denote the result by Degree, BP, and SP Non-Preserving, respectively.}
\label{s9-1}
\end{figure}

\begin{figure}
{\LARGE OpenFlights}
\centering
\includegraphics[width=\textwidth]{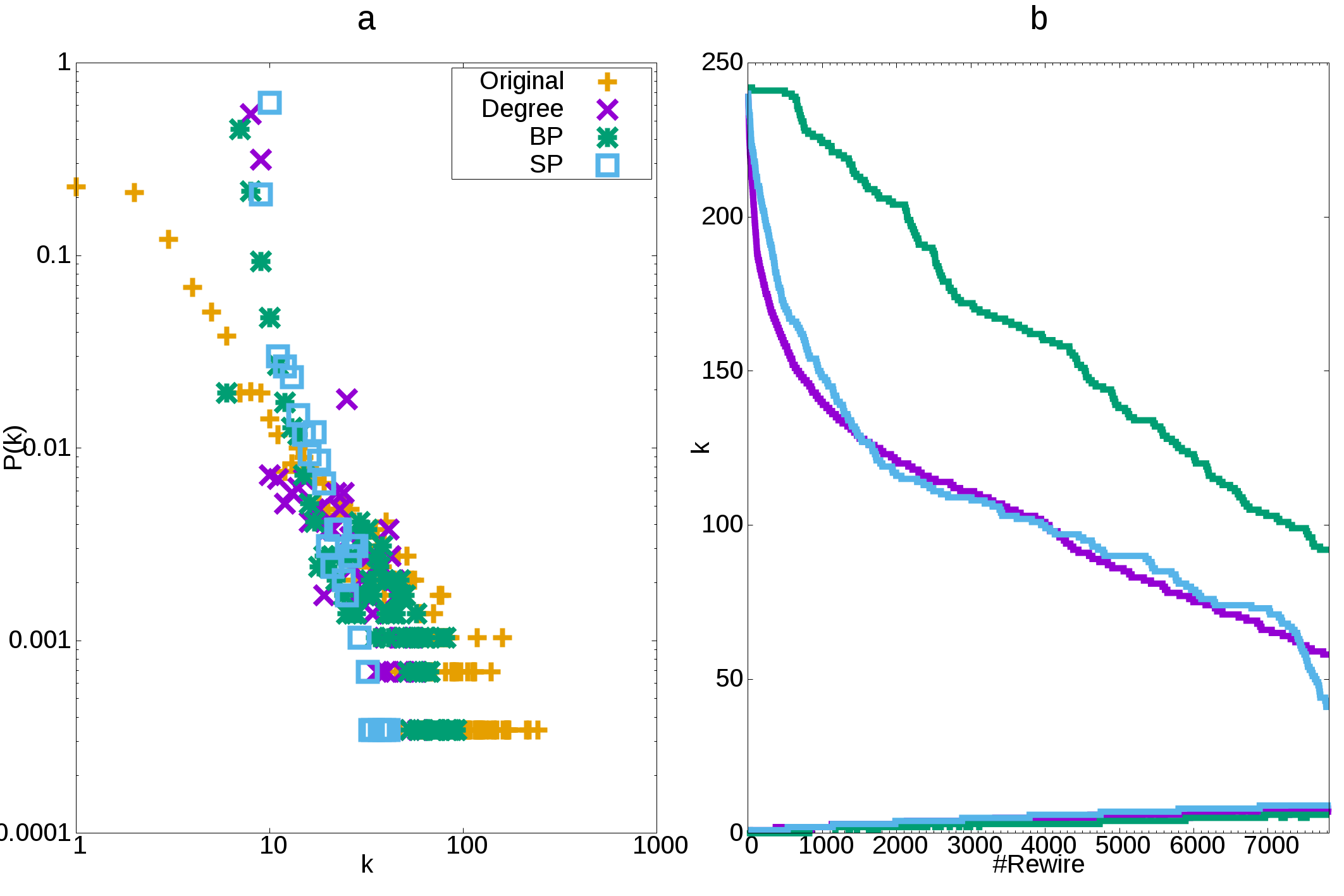}
\caption{OpenFlights \cite{kunegis2013konect,opsahl2010node}. (a) Degree distributions in original and after rewiring networks, (b) Maximum and minimum degrees vs. the number of rewiring in Degree, BP, and SP Non-Preserving. The above three lines show the maximum degrees. The below three lines show the minimum degrees. Violet, green, and light blue denote Degree, BP, SP Non-Preserving. Orange denotes the original degree distribution.}
\label{s9-2}
\end{figure}

\begin{figure}
{\LARGE PolBlogs}
\centering
\includegraphics[width=\textwidth]{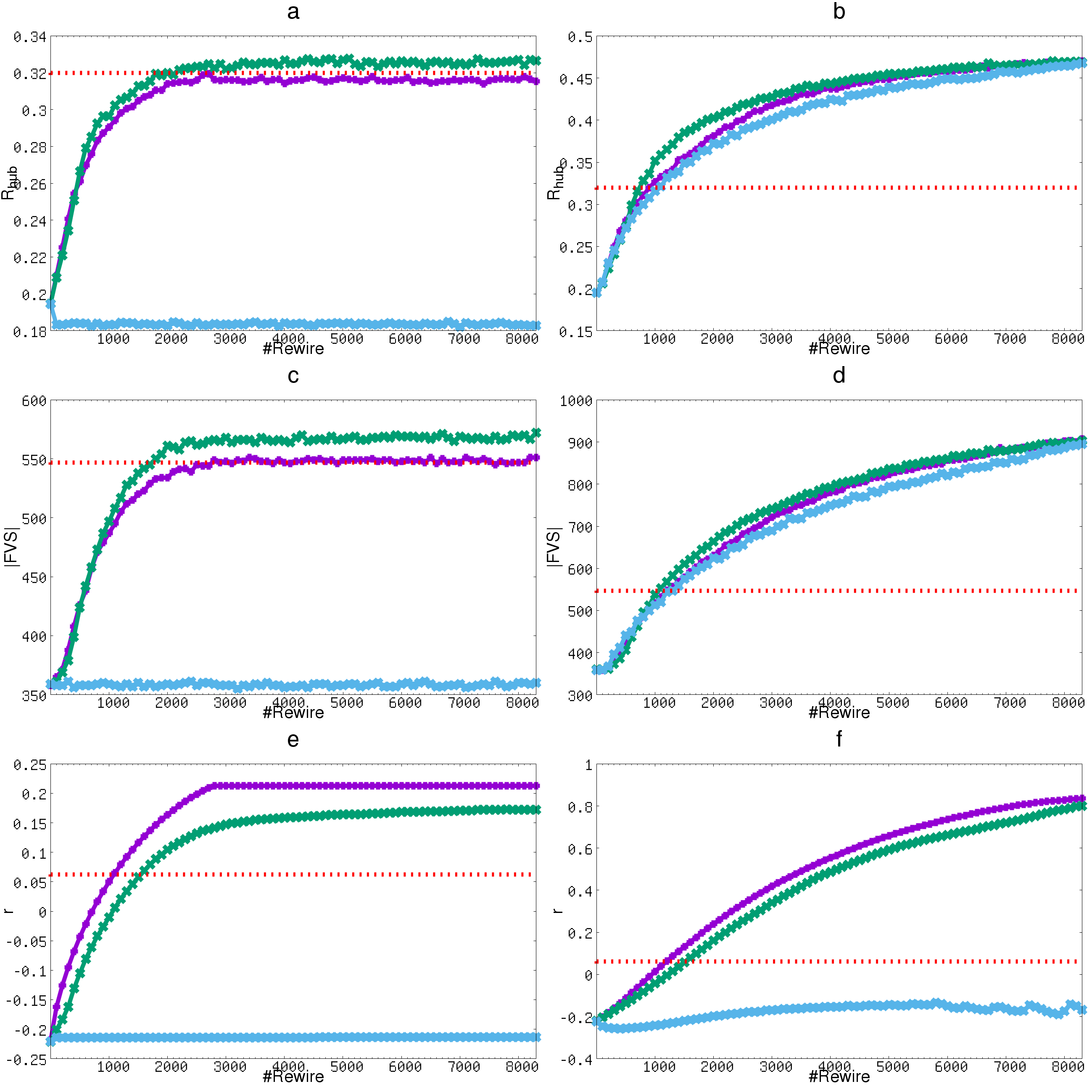}
\caption{PolBlogs \cite{adamic2005political}. Comparison of the robustness index $R_{\mathrm{hub}}$, the size approximate of FVS, and the degree-degree correlation coefficient $r$ vs. the number of rewiring. (Left: a, c, e) Rewirings in Preserving. Violet, green, and light blue solid lines denote the result by Degree, BP, and SP Preserving, respectively. The red dot line indicates a baseline of the conventional best. (Right: b, d, f) Rewirings in Non-Preserving. Violet, green, and light blue solid lines denote the result by Degree, BP, and SP Non-Preserving, respectively.}
\label{s10-1}
\end{figure}

\begin{figure}
{\LARGE PolBlogs}
\centering
\includegraphics[width=\textwidth]{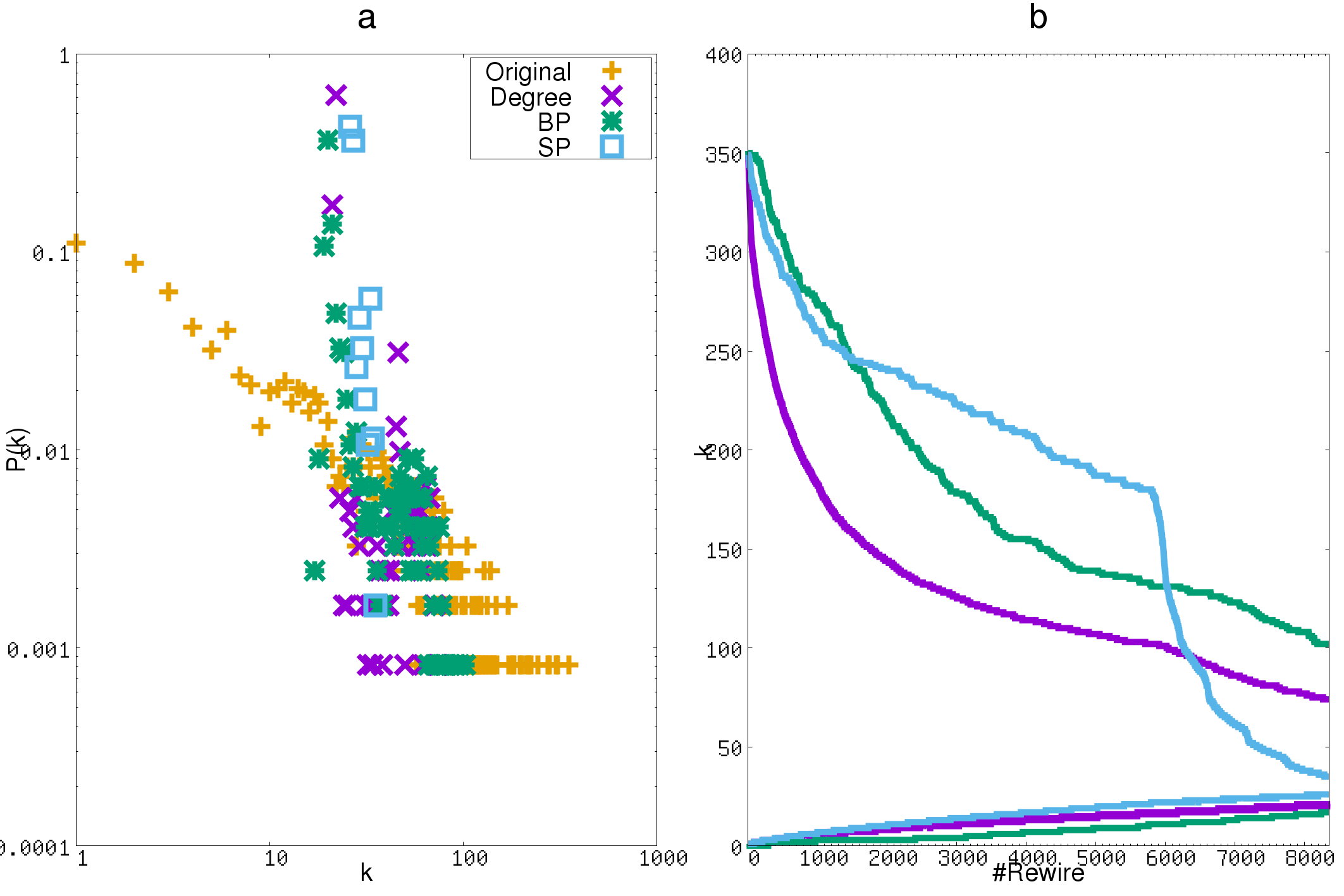}
\caption{PolBlogs \cite{adamic2005political}. (a) Degree distributions in original and after rewiring networks, (b) Maximum and minimum degrees vs. the number of rewiring in Degree, BP, and SP Non-Preserving. The above three lines show the maximum degrees. The below three lines show the minimum degrees. Violet, green, and light blue denote Degree, BP, SP Non-Preserving. Orange denotes the original degree distribution.}
\label{s10-2}
\end{figure}

\clearpage
\bibliographystyle{bmc-mathphys}
\bibliography{supplement}